\documentclass[12pt]{article}
\usepackage[margin=1in]{geometry}
\usepackage{setspace}
\usepackage{color}
\usepackage[T1]{fontenc}
\usepackage{lmodern}
\usepackage{amsmath,amssymb,amsthm,mathtools,booktabs,array}
\usepackage{microtype}
\usepackage[colorlinks=true,linkcolor=blue,citecolor=blue]{hyperref}
\usepackage{enumitem}
\setlist[itemize]{leftmargin=*,itemsep=2pt,topsep=3pt}
\allowdisplaybreaks[2]
\numberwithin{equation}{section}
\newtheorem{theorem}{Theorem}[section]
\newtheorem{definition}{Definition}[section]
\newtheorem{proposition}[theorem]{Proposition}
\newtheorem{lemma}{Lemma}[section]
\newtheorem{remark}{Remark}[section]
\newtheorem{example}{Example}[section]
\newcommand{\E}{\mathbb E}
\newcommand{\PP}{\mathbb P}

\newcommand{\one}{\mathbf 1}
\newcommand{\I}{(0,1)}
\newcommand{\Dp}{D_+}
\newcommand{\Dm}{D_-}
\newcommand{\Ds}{D_*}
\newcommand{\DG}{D_G}
\newcommand{\SG}{S_G}

\title{Smooth Truth Serum Mechanisms: Two-Player Characterization, Equilibrium Non-Domination, and Information Sensitivity
	}
\author{Jak\v{s}a Cvitani\'c\thanks{Division of the Humanities and Social Sciences, Caltech. E-mail: cvitanic@hss.caltech.edu.}
	\and Dra\v{z}en Prelec\thanks{MIT, Sloan School of Management, Department of Economics, Department of Brain and Cognitive Sciences. E-mail: dprelec@mit.edu. Thanks are due to All Souls College, Oxford, for Visiting Fellowships during Michaelmas 2020, Hillary 2021, and Michaelmas 2022 terms.}
	\and Sonja Radas\thanks{The Institute of Economics, Zagreb, and MIT, Sloan School of Management. E-mail: sradas@eizg.hr; sradas@mit.edu. This work was made as part of the project ``Determinants of Strengthening Technological Capabilities of Different Sectors'' at the Institute of Economics, Zagreb and funded within the National Recovery and Resilience Plan 2021--2026 -- NextGenerationEU.}
	\and Hrvoje \v{S}iki\'c\thanks{University of Zagreb, Faculty of Science, Department of Mathematics. E-mail: hsikic@math.hr. Research supported in part by the Croatian Science Foundation under project HRZZ-IP-2022-10-5116 (FANAP). Hrvoje \v{S}iki\'c was also supported by the European Union -- NextGenerationEU through the National Recovery and Resilience Plan 2021--2026 Institutional grant of University of Zagreb, Faculty of Science (IK IA 1.1.3. Impact4Math).}}
\date{November 25, 2026}

\begin{document}
	\maketitle
	\begin{abstract}
		We construct mechanisms of Bayesian Truth Serum (BTS) type for eliciting honest answers and predictions on the distribution of the answers that are: (i) incentive compatible for honesty (IC), (ii) ex-ante non-dominated (END) by dishonest equilibria. In the case of $m=2$ answer choices   and $n=2$ respondents, we completely characterize continuously differentiable IC mechanisms, we find necessary conditions for END and construct an IC END example.
		 The reward separates into three terms: (1) a proper score that provides incentives for honest prediction; (2)  a penalty for matching answers with inconsistent predictions; and (3) an arbitrary term that depends only on the other respondent's reports.  The pairwise construction extends IC and END to arbitrary finite numbers of respondents and answers. The infinite-population limit is also IC.
		 For positively correlated binary types and two respondents, we obtain information-sensitive honest payments that decrease under garbling and conditions for implementing prescribed interim payoffs. 

	\end{abstract}
	
	\section{Introduction}

Mechanisms of the Bayesian truth serum (BTS) kind reward honest answers to questions about beliefs, behaviors, preferences or intentions, as might be elicited in a social science survey or lab experiment (Prelec, 2004). Respondents report an answer, where the honest answer is their {\it{type}}, to a multiple choice question and a prediction on the distribution of answers. A scoring rule mechanism computes a payment for each respondent from these reports. Respondents are assumed to care only about their (expected) payments. A mechanism is incentive compatible (IC) if honest answers and honest predictions form a Bayesian equilibrium, and strictly IC if honesty is strictly better than dishonesty against honest peers.
We call an IC mechanism ex-ante non-dominated (END) if no dishonest finite-payoff Bayesian equilibrium gives both respondents at least their honest ex-ante expected payments and one of them strictly more. 

The simplest case has $n=2$ respondents and $m=2$ possible answers, which is our building block for the general case. The two-respondent problem arises naturally when two readers assess the same image, two referees (or AI models) evaluate the same project, two doctors give a medical opinion, and so on.
Even in this case, strict IC cannot be guaranteed for all joint distributions when only answers and predictions are elicited (Prelec, 2021). Rather than asking respondents for additional information, we assume that their types are exchangeable.
 Our main results address three economic questions.

{\bf First, which payment rules make honesty an equilibrium?}
 For two respondents and binary answers, Theorem~\ref{thm:main} characterizes all
 continuously differentiable IC mechanisms on the full domain of exchangeable,
 stochastically relevant type distributions. Every such payment rule has three
 components: a proper scoring rule that rewards accurate predictions, a term
 depending only on the other respondent's report, and a penalty
 imposed when the respondents give the same answer but report different
 predictions. 
 Theorem~\ref{thm:multi-rep} extends this characterization to two respondents with an
 arbitrary number of possible answers, under an additional condition on
 payments near the vertices of the probability simplex.
 
{\bf Second, how can honesty be protected against more profitable equilibria?}
 For two respondents receiving payments from the same rule, Theorem~\ref{thm:END-normal}
 establishes necessary conditions for END. In particular, END on the full domain  requires the honest ex-ante expected payment to be
 a constant, independent of the joint distribution of types. It also
 restricts the component of the payment that depends only on the peer's
 report.
 We then construct a family of strictly IC and END mechanisms using weighted
 Bregman-divergence penalties (Theorem~\ref{thm:construct}). For these mechanisms, every
 Bayesian equilibrium satisfies
 $
 \mathbb E[F_1+F_2]\le 2C,
 $
 where $C$ is each respondent's honest ex-ante expected payment. This
 aggregate bound rules out any equilibrium that Pareto-dominates honesty,
 including asymmetric and mixed equilibria.
 
 {\bf Third, which results extend to larger populations and more possible answers?}
 Theorem~\ref{thm:multi-end} shows that for $n\ge2$ respondents and $m\ge2$
 possible answers, averaging suitable pairwise $m$-answer scores preserves IC
 under the stated exchangeability and stochastic-relevance assumptions.
 Its weighted Bregman construction is strictly IC and END for every finite $n$.
 In particular, every Bayesian equilibrium satisfies
 $
 \sum_{i=1}^n\mathbb E[F_i]\le nC,
 $
 while honesty gives every respondent expected payment $C$. 
 We also consider infinitely many respondents. 
 Although the original Bayesian Truth Serum also has
 non-domination properties with an infinite population, our construction yields an
 explicit aggregate equilibrium bound for every finite population and permits
 bounded payments for suitable scoring functions.
 
 {\bf Additional results: rewarding information.}
 When the two binary types  are positively
 correlated, more flexibility is possible.
 Let $x$ denote the probability that the other respondent's type is one
 conditional on one's own type being one, and $y$ the corresponding
 probability conditional on one's own type being zero. On the domain $x>y$,
corresponding to positive correlation and consensus bias, we construct a
 strictly IC and END mechanism whose honest expected payment decreases when
 the same noisy binary channel is applied independently to both type signals
 (Theorems~\ref{thm:equilibrium} and~\ref{thm:garbling}). Thus, honest payments can reflect the informativeness
 of the respondents' answers.
 We also consider the implementation of prescribed interim expected payments.
 Under a restriction keeping beliefs and reported predictions away from zero
 and one, Theorem~\ref{P:pair} gives sufficient conditions for implementing such
 payments by an IC mechanism. One example rewards information gain, measured
 by the Kullback--Leibler divergence between a respondent's posterior and
 prior beliefs (Example~\ref{ex:information-gain}).


 {\bf Literature.} 
Many IC mechanisms require more than two respondents. The original Bayesian truth serum requires a sufficiently large population
 (Prelec, 2004). The robust Bayesian truth serum of Witkowski and Parkes (2012) for binary types requires at least three respondents. Bayesian markets (Baillon, 2017) and binary choice-matching (Cvitani\'c et al., 2019) also need more than two respondents.
The mechanism of Radanovic and Faltings (2013) works with two respondents but requires $x>y$ in the binary case. Their divergence-based truth serum (Radanovic and Faltings, 2014) also uses a prediction score and penalizes reports whose answers agree while their predictions are inconsistent.
Papers that identify necessary conditions include: Frongillo and Kash (2021) develop
a general framework for truthful mechanisms, while Frongillo and Witkowski (2017)
characterize minimal peer-prediction mechanisms geometrically. Our characterization differs in that it
concerns the joint report of an answer and a prediction.
Mechanisms  that have non-domination properties (and typically ask  different  questions than BTS) include Kong and Schoenebeck (2018), 
Schoenebeck and Yu (2023),
Shnayder et al. (2016),
Kong and Schoenebeck (2019) and  Kong (2024).
Equilibrium robustness has also been studied through coalitional deviations.
Han, Schoenebeck, Tao and Xia (2025) analyze Bayesian equilibria robust to
deviations by groups of bounded size. This criterion differs from  our END. Robust decentralized verification without ground truth is studied in Zhao, Chen and Zhou (2025).
Related, but different from ours approaches to information incentives include the stochastic-dominance
truthfulness criterion of Zhang et al. (2025) and the finite-sample information-measure
analysis of Aznag et al. (2026). In another recent paper, Peker (2025) develops peer-prediction mechanisms for aggregating correlated expert judgments.
  Lehmann (2026) surveys twenty-five mechanisms for information elicitation without verification and highlights the problems posed by nontruthful equilibria, including those yielding higher expected rewards than honesty. 
The current paper also builds on our earlier work on choice-matching (Cvitani\'c et al., 2019) and IC surveys based on posterior probabilities (Cvitani\'c et al., 2020). Unlike the existing literature, we focus on characterizing smooth IC mechanisms and on the ex-ante non-domination of honesty by alternative Bayesian equilibria.


Section~\ref{sec:model} presents the model, Section~\ref{sec:normal} gives the characterization in the $2\times2$ case, and Section~\ref{sec:non-domination} studies non-domination. Section~\ref{sec:many} considers the general $m\times n$ case, including infinitely many respondents. Section~\ref{sec:garbling} presents the additional results for $x>y$.
Proofs are in the appendix.

\section{Model with two respondents and two choices}\label{sec:model}
	Two respondents are assigned exchangeable binary types  $T_i\in\{0,1\}$. Write
	\begin{equation}\label{eq:beliefs}
		x=\PP(T_{-i}=1\mid T_i=1),\qquad
		y=\PP(T_{-i}=1\mid T_i=0)
		,\qquad p_1=x,\quad p_0=y
		.
	\end{equation}
	We denote
	\begin{equation}\label{eq:domains}
		\Dp=\{0<y<x<1\},\qquad \Dm=\{0<x<y<1\},\qquad \Ds=\Dp\cup\Dm.
	\end{equation}

	A {\bf report} is  a pair $(j,z)\in\{0,1\}\times\I$: $j$ is the {\bf reported answer} (which need not equal the respondent's type) and $z$ is a {\bf prediction} of the probability  that the peer's type is $1$. Against peer report $(k,u)$, the reporting respondent $i$'s payment is given by reward functions  $f^i_{jk}(z,u)$. 
	The planner does not know $(x,y)$, which is why these functions do not depend on it. The respondents know $(x,y)$, and the joint distribution is common knowledge among them. The honest report is $(T_i,p_{T_i})$.

	Respondents may use asymmetric mixed reporting strategies, with private
	randomizations independent conditional on the types. Write $(J_i,Z_i)$
	for respondent $i$'s report and
	$F_i=f^i_{J_iJ_{-i}}(Z_i,Z_{-i})$ for the realized payment.

	We say that a mechanism corresponding to reward functions $f^i_{jk}(z,u)$ is
	{\bf incentive compatible} ({\bf IC}) on $D$ if honesty is a Bayesian equilibrium
	for every $(x,y)\in D$. It is {\bf strictly IC} on $D$ if, for every
	$(x,y)\in D$, $i\in\{1,2\}$, $t\in\{0,1\}$, and
	$(j,z)\ne(t,p_t)$,
	\[
	\E\!\left[f^i_{t,T_{-i}}(p_t,p_{T_{-i}})\mid T_i=t\right]
	>
	\E\!\left[f^i_{j,T_{-i}}(z,p_{T_{-i}})\mid T_i=t\right].
	\]
	We use abbreviation IC also for {\bf incentive compatibility}.

	\begin{remark}\ 
		
		(i) While exchangeability is not necessary, Prelec (2021) gives non-exchangeable examples for which strict IC cannot be achieved with only the answer and prediction reports. 
		
		(ii) If $x=y$, the two types have the same conditional belief, and strict IC is impossible: against a truthful peer, each type faces the same distribution of peer reports, so their distinct truthful reports cannot both be uniquely optimal.  
		
		(iii) Some results can only be obtained on a restricted domain, $\Dp$ or $\Dm$. The domain $\Dp$ is  the positive-correlation case:
$\operatorname{Cov}(T_1,T_2)=\pi(1-\pi)(x-y)>0$, where
$\pi=\PP(T_i=1)$. If the types belong to an infinitely
exchangeable sequence, by de Finetti's theorem there exists a ``state of nature" $\Theta$ such that
$\operatorname{Cov}(T_1,T_2)=\operatorname{Var}(\Theta)\ge0$,
with $x>y$ exactly when $\operatorname{Var}(\Theta)>0$.
The frequently invoked ``consensus bias'' is an informal interpretation
of this positive-correlation restriction.

		(iv)
		%
		The interior restrictions $x,y\in(0,1)$ ensure that the conditional beliefs and derivatives used below are well defined. Moreover, it can be shown that if we allow $x, y \in [0, 1]$ instead (with $x\neq y$), then there is no
		strictly IC mechanism continuous on $[0,1]$. 
	\end{remark}


	\section{Characterization of \texorpdfstring{${\bf C^1}$}{C1} IC mechanisms}\label{sec:normal}
	
	For a given convex function $G\in C^1(\I)$ 
	define the proper score $S_G$ by
	\begin{equation}\label{eq:score}
		\SG(z,k)=G(z)+(k-z)G'(z).
	\end{equation}
	
	\begin{theorem}
		\label{thm:main}
		The mechanism determined by $f^i_{jk}\in C^1(\I^2)$ is IC on $\Ds$ if and only if there exist  convex functions $G^i\in C^2(\I)$, functions $B^i_0,B^i_1\in C^1(\I)$, and $K^i_0,K^i_1\in C^1(\I^2)$ such that
		\begin{equation}\label{eq:normal-form}
			\boxed{f^i_{jk}(z,u)=\SG^i(z,k)+B^i_k(u)-\one_{\{j=k\}}K^i_j(z,u),}
		\end{equation}
		with
		\begin{equation}\label{eq:penalty-conds}
			K^i_j(z,u)\ge0,\qquad K^i_j(u,u)=0\quad(j=0,1).
		\end{equation}
		It is strictly IC on $\Ds$
		if and only if $G^i$ are strictly convex and $K^i_j(z,u)>0$ whenever $z\ne u$.
	\end{theorem}
	
\noindent The proof is in Appendix~\ref{app:thm31}, and here is its summary: for necessity,
truthful prediction first-order conditions give off-diagonal separability of the form $S_G+B$;
The fact that honesty has to dominate the strategy of reporting the other type's truthful report implies  a common convex
scoring rule $S_G$. Deviating only in the reported answer yields the
nonnegative matching penalties $K^i_j(z,u)$, subtracted when the reported
answers are equal $j$, and equal to zero if the reported predictions coincide. Sufficiency and strictness follow from
$S_G$ being a proper scoring rule and the form of the penalty term.

\begin{remark}[Affine non-uniqueness]
The representation \eqref{eq:normal-form} is not unique. For any
affine function $\ell(z)=az+b$,
\[
S_{G+\ell}(z,k)=S_G(z,k)+ak+b.
\]
For each respondent $i$, replacing
$(G^i,B^i_0,B^i_1)$ by
$(G^i+\ell,B^i_0-b,B^i_1-a-b)$ leaves
\eqref{eq:normal-form} unchanged, with $K^i_j$ fixed.
\end{remark}

	  \subsection{Mechanisms which are not \texorpdfstring{$C^1$}{C1}}
	  There are many IC mechanisms in the case $m=n=2$ which are not $C^1$, or even continuous, but we do
	  not have a clean characterization of those. 
	  
	  {\bf A separable family.} We provide a general sufficient
	  condition for IC, which mimics the structure of the characterization of $C^1$
	  mechanisms in Theorem~\ref{thm:main}.
	  Take any two functions $B_0(u),B_1(u)$, any two functions $K_0(z,u),K_1(z,u)$
	  such that
	  \[
	  K_j(u,u)=0,\qquad K_j(z,u)>0\quad\text{for }z\ne u,
	  \]
	  and any two functions $H_0(x),H_1(x)$ such that, for $z\ne u$,
	  \begin{equation}\label{eq:nonc1}
	  	\bigl(H_1(z)-H_1(u)\bigr)z
	  	+\bigl(H_0(z)-H_0(u)\bigr)(1-z)\ge0.
	  \end{equation}
	  For example, this holds for  $H_k(z)=S(z,k)$, $k=0,1$, where $S$ is a proper scoring rule. Define
	  \[
	  f_{jk}(z,u)=H_k(z)+B_k(u)-\one_{\{j=k\}}K_j(z,u).
	  \]
	  Then this is an IC mechanism on $\Ds$. It is strictly IC if the inequality in
	  \eqref{eq:nonc1} is strict. The proof is obtained by direct verification and
	  is similar to the proof of sufficiency in Theorem~\ref{thm:main}.
	  
	{{\bf A continuous non-separable  IC mechanism.}}
	  	Define
	  	\[
	  	g(z):=(1-z^2)^2,\qquad
	  	L(z,u):=\sum_{p\in\{1,2,4\}}|z^p-u^p|.
	  	\]
	  	Consider the mechanism given by
	  	\begin{equation}
	  		\boxed{\displaystyle
	  			f_{jk}(z,u)=-4\mathbf1_{\{j=k\}}L(z,u)
	  			+\mathbf1_{\{j\ne k\}}(2j-1)
	  			\bigl[ju+(1-j)(1-u)\bigr]\bigl[g(z)-g(u)\bigr].}
	  		\label{eq:continuous-example}
	  	\end{equation}
	  	As proved in Appendix~\ref{app:nonc1}, this mechanism is strictly IC on $\Ds$, and it is bounded and continuous. It shows that continuity does not imply the representation of  Theorem \ref{thm:main}
	  	without its $C^1$ assumption.

	\section{Necessary conditions for non-domination and an example}\label{sec:non-domination}

	IC does not exclude other equilibria that both respondents may prefer to honesty.
	We now study what is needed to rule out such Pareto improvements.

\begin{definition}[Ex-ante non-domination on a domain]
Fix $D\subseteq\Ds$ and an IC mechanism on $D$.
A Bayesian equilibrium is \emph{finite-payoff} if
$\E[F_i\mid T_i=t]\in\mathbb R$ for $i=1,2$ and $t=0,1$.
Honesty is \emph{ex-ante non-dominated} (END) on $D$ if, for
\emph{every} $(x,y)\in D$, there is no other finite-payoff Bayesian
equilibrium $\tau$ such that
\[
\E[F_i^\tau]\ge\E[F_i^H]\quad(i=1,2),
\qquad \E[F_i^\tau]>\E[F_i^H]\quad\text{for at least one }i.
\]
The definition includes asymmetric and mixed equilibria.
\end{definition}
	
	Next, we derive necessary conditions for END and give an example of  a family of
	END mechanisms. We do not obtain a complete characterization of END.
	
	For $r\in(0,1)$, let $U_r$ be the strategy profile in which each
respondent reports answer $1$ with probability $r$
(independently of the other respondent), and always reports prediction $r$.
These profiles are equilibria of every IC mechanism considered in this section,
as shown in Appendix~\ref{app:thm41}.

\begin{theorem}\label{thm:END-normal}
		Consider an IC mechanism on $\Ds$ such that, for every
$(x,y)\in\Ds$, the honest equilibrium is not ex-ante dominated
by any profile $U_r$, $r\in(0,1)$. Assume 
		that both respondents are paid by the same four functions $f_{jk}\in C^1(\I^2)$.
		Then, 
		there exist $C\in\mathbb R$, a convex function $\widehat G\in C^2(\I)$, and functions $K_0,K_1\in C^1(\I^2)$ with $K_j\ge0$ and $K_j(u,u)=0$, such that
		\begin{equation}\label{eq:ENDnormal}
			\boxed{f_{jk}(z,u)=C+S_{\widehat G}(z,k)-\widehat G(u)-\one_{\{j=k\}}K_j(z,u).}
		\end{equation}
		Furthermore, each respondent obtains constant ex-ante expected payment $C$ under honesty, for all $(x,y)\in \Ds$. 
	\end{theorem}	
	
The proof is in Appendix~\ref{app:thm41}.
It shows that: the common expected payment $R(r)$ under $U_r$ is constant
in $r$; non-domination by these profiles results in constant honest
ex-ante payments and $B_k(u)=C-G(u)+a_0(u-k)$, for $k=0,1$ and some
$a_0\in\mathbb R$; then, the generator can be normalized to
$\widehat G(u)=G(u)-a_0u$, yielding \eqref{eq:ENDnormal}.


	\subsection{Example of an END mechanism }\label{sec:constructive}
	Assume $G$ extends continuously to $[0,1]$, belongs to $C^2(\I)$, and satisfies $G''(z)>0$ throughout $\I$.
	Define the Bregman divergence by
	\begin{equation}\label{eq:bregman}
		\DG(p,z)=G(p)-G(z)-(p-z)G'(z).
	\end{equation}
	For a respondent of type $t$, let $p=p_t=\PP(T_{-i}=1\mid T_i=t)$.
	Against a truthful peer, $k=T_{-i}$, and the difference between truthful
	and untruthful expected scores is the Bregman divergence:
	\begin{equation}\label{BD}
		\mathbb{E}_p[S_G(p,k)]-\mathbb{E}_p[S_G(z,k)]
		=D_G(p,z).
	\end{equation}
	Let
	\begin{equation}\label{eq:weights}
		\mu=\min_{u\in[0,1]}G(u),\qquad
		w_j(u)=\kappa_j+\lambda_j[G(u)-\mu],\qquad \kappa_j>0,\quad\lambda_j\ge0.
	\end{equation}
	Assume that both respondents use the common payment functions $f_{jk}$.
	
	\begin{theorem}[Non-domination by every  equilibrium]\label{thm:construct}
		The mechanism
		\begin{equation}\label{eq:family}
			\boxed{f_{jk}(z,u)=C+\SG(z,k)-G(u)-\one_{\{j=k\}}w_j(u)\DG(u,z)}
		\end{equation}
		is strictly IC and satisfies END on $\Ds$. Every equilibrium (including asymmetric and/or mixed equilibria) satisfies
		\begin{equation}\label{eq:budget}
			\boxed{\E[F_1+F_2]
				=2C-\E\!\left[\one_{\{J_1=J_2\}}\lambda_{J_1}
				\bigl(G(Z_1)-G(Z_2)\bigr)^2\right]\le2C.}
		\end{equation}
		Moreover, each respondent's honest ex-ante payment is $C$.
	\end{theorem}
	
This result is a special case of	Theorem~\ref{thm:multi-end} below: take
$m=n=2$ and $w_j(u)=\kappa_j+\lambda_j[G(u)-\mu]$.
Then \eqref{eq:multi-all-to-all-bound}
reduces to \eqref{eq:budget}; its strict-IC and END conclusions
and the honest payment $C$ follow as well. 
	
\begin{remark}	
	\ 
	
	(i) The result is also true more generally for the weights of the form $w_j(u)=\psi_j(G(u))$ for positive, continuous, nondecreasing functions $\psi_j$ on $G([0,1])$; see Theorem~\ref{thm:multi-end}. 
	
	 (ii) Note that constant weights give $\E[F_1+F_2]=2C$ in every equilibrium. 
	
	 (iii) Section~\ref{sec:garbling} gives an END IC mechanism on $\Dp$ with nonconstant honest ex-ante payments.
	\end{remark}

\section{Many types  and many respondents}\label{sec:many}

\subsection{Two respondents and \texorpdfstring{$m\ge2$}{m>=2} types}\label{sec:multi-signals}
We extend the two-respondent characterization to $m\ge2$ possible types. Let
\begin{equation}\label{eq:multi-simplex}
 \Delta_m^\circ:=\left\{z\in(0,1)^m:\sum_{k=1}^m z_k=1\right\},
 \qquad
 S_m:=\left\{v\in\mathbb R^m:\sum_{k=1}^m v_k=0\right\},
\end{equation}
and let $e_k\in\mathbb R^m$ be the $k$th coordinate vector. For a
symmetric joint distribution $P=(P_{st})_{s,t=1}^m$, write
\begin{equation}\label{eq:multi-rows}
 P_{t\cdot}:=(P_{t1},\ldots,P_{tm}),\qquad
 \pi_t(P):=\sum_{k=1}^m P_{tk},\qquad
 p_t(P):=\frac{P_{t\cdot}}{\pi_t(P)}\in\Delta_m^\circ.
\end{equation}
For $n=2$, the two respondents observe private types $T_1,T_2\in\{1,\ldots,m\}$, whose joint distribution is represented by the matrix $P=(P_{jk})_{j,k=1}^m$. The following is the set of admissible joint type distributions:
\begin{equation}\label{eq:multi-domain}
 \mathcal E_m:=\left\{
 P\in(0,1)^{m\times m}:
 P=P^{\mathsf T},\quad \sum_{s,t=1}^m P_{st}=1,\quad
 p_s(P)\ne p_t(P)\ \text{for all }s\ne t
 \right\}.
\end{equation}
Each respondent observes $T_i\in\{1,\ldots,m\}$ and reports
$(j,z)\in\{1,\ldots,m\}\times\Delta_m^\circ$.\footnote{For $m=n=2$, identify the binary answers $1,0$ with $e_1,e_2$, respectively, and the scalar prediction $z\in(0,1)$ with $(z,1-z)\in\Delta_2^\circ$. Given the scalar generator $G$, define the simplex generator $\bar G(z_1,z_2)=G(z_1)$. Then, for $S_{\bar G}$ and $D_{\bar G}$ as defined in the following theorem, we have  $S_{\bar G}((z,1-z),1)=S_G(z,1)$, $S_{\bar G}((z,1-z),2)=S_G(z,0)$, and $D_{\bar G}((u,1-u),(z,1-z))=D_G(u,z)$.}
	The honest report of type $t$
is $(t,p_t(P))$. Against a truthful peer, this report is optimal if
\begin{equation}\label{eq:multi-ic}
 \sum_{k=1}^m p_t(P)_k f^i_{tk}(p_t(P),p_k(P))
 \ge \sum_{k=1}^m p_t(P)_k f^i_{jk}(z,p_k(P))
\end{equation}
for every $P\in\mathcal E_m$, $t,j\in\{1,\ldots,m\}$,
$z\in\Delta_m^\circ$, and respondent $i=1,2$. Strictly IC means strict
inequality in \eqref{eq:multi-ic} whenever $(j,z)\ne(t,p_t(P))$.

Derivatives on $\Delta_m^\circ$ are relative to its affine hull:
$C^r(\Delta_m^\circ)$ means $C^r$ in coordinates
$z_1,\ldots,z_{m-1}$, with $z_m=1-\sum_{\ell<m}z_\ell$. Equivalently,
for $v\in S_m$,
\begin{equation}\label{eq:multi-deriv}
 \mathrm D G(z)[v]
 :=\left.\frac{d}{dh}G(z+hv)\right|_{h=0},\qquad
 \mathrm D_1 f_{jk}(z,u)[v]
 :=\left.\frac{d}{dh}f_{jk}(z+hv,u)\right|_{h=0}.
\end{equation}
In particular $e_k-z\in S_m$; thus $\mathrm D G(z)[e_k-z]$
is well defined without extending $G$ beyond the simplex.

The following {\it vertex} condition requires each difference between off-diagonal payments with peer answer $k$ to have a finite, path-independent limit as the peer prediction approaches $e_k$. We impose it for every $k$, every
$a,b\ne k$ (including $a=b$), and every $z,z'\in\Delta_m^\circ$:
\begingroup
\renewcommand{\theHequation}{vertexV}
\begin{equation}\tag{V}\label{eq:multi-vertex}
 \lim_{\substack{u\to e_k\\u\in\Delta_m^\circ}}
 \bigl[f^i_{ak}(z,u)-f^i_{bk}(z',u)\bigr]\in\mathbb R,
 \qquad i=1,2.
\end{equation}
\endgroup

\begin{theorem}[Characterization for two respondents]\label{thm:multi-rep}
	Let $m\ge2$, $n=2$. Suppose $f^i_{jk}\in C^1(\Delta_m^\circ\times\Delta_m^\circ)$.
	The mechanism is IC on $\mathcal E_m$ and satisfies
	\eqref{eq:multi-vertex} if and only if, for $i=1,2$,
	\begin{equation}\label{eq:multi-normal}
		\boxed{
			f^i_{jk}(z,u)=S_{G^i}(z,k)+B^i_k(u)
			-\one_{\{j=k\}}K^i_j(z,u),\qquad
			S_{G^i}(z,k)=G^i(z)+\mathrm D G^i(z)[e_k-z],
		}
	\end{equation}
	where $G^i\in C^2(\Delta_m^\circ)$ is convex,
	$B^i_k\in C^1(\Delta_m^\circ)$, and
	$K^i_j\in C^1(\Delta_m^\circ\times\Delta_m^\circ)$, with
	\begin{equation}\label{eq:multi-K}
		K^i_j(z,u)\ge0,\qquad K^i_j(u,u)=0.
	\end{equation}
	Within this class, strict IC is equivalent to strict
	convexity of every $G^i$ and $K^i_j(z,u)>0$ whenever $z\ne u$.
\end{theorem}
The proof is in Appendix~\ref{app:multi-rep}. For $m=2$, the result is Theorem~\ref{thm:main} under the binary-to-simplex identification above, and condition~\eqref{eq:multi-vertex} is automatic.

\subsection{General \texorpdfstring{$m\ge2$}{m>=2} and \texorpdfstring{$n\ge2$}{n>=2}: IC and END}\label{sec:multi-general}
For $n\ge2$, let $(T_1,\ldots,T_n)$ have an exchangeable
joint distribution with common two-respondent marginal $P\in\mathcal E_m$.
Each respondent $i$ reports $(J_i,Z_i)$ and receives the average
of the scores against the other $n-1$ respondents:
\begin{equation}\label{eq:multi-all-to-all}
\begin{aligned}
 F_i^{(n)}:=\frac1{n-1}\sum_{\ell\ne i} f^i_{J_iJ_\ell}(Z_i,Z_\ell),
 \qquad i=1,\ldots,n.
\end{aligned}
\end{equation}
For common payment functions $f_{jk}$, the superscript $i$ is omitted.
For $n$ respondents, IC means that truthful reporting is a Bayesian
equilibrium of the game with payments \eqref{eq:multi-all-to-all};
strictly IC means honesty is the unique best response for each type against honest
opponents. END means that, for every admissible exchangeable joint law,
no other Bayesian equilibrium with finite payoffs $\E[F_i^{(n)}\mid T_i=t]$
for every $i,t$ weakly improves all $n$ respondents' ex-ante payments,
with a strict improvement for at least one.

\begin{theorem}[IC and END for arbitrary $m,n\ge2$]\label{thm:multi-end}
For every $n\ge2$, the representation \eqref{eq:multi-normal}
	(with $i=1,\ldots,n$) is also sufficient for IC of the average
	mechanism \eqref{eq:multi-all-to-all}; strict convexity and
	strictly positive matching penalties suffice for strict IC.

	Let $m\ge2$ and $n\ge2$, and suppose all respondents use the same pairwise
	payment functions. Let
	$G\in C(\Delta_m)\cap C^1(\Delta_m^\circ)$ be strictly convex on $\Delta_m^\circ$,
	where $\Delta_m$ is the closed probability simplex. Let
	$\psi_j>0$ be continuous and nondecreasing on $G(\Delta_m)$,
	and define
	$w_j(u):=\psi_j(G(u))$ and
	$D_G(u,z):=G(u)-G(z)-\mathrm DG(z)[u-z]$.
	Then the common pairwise score
	\begin{equation}\label{eq:multi-end-mechanism}
		\boxed{f_{jk}(z,u)=C+S_G(z,k)-G(u)
			-\one_{\{j=k\}}w_j(u)D_G(u,z)}
	\end{equation}
	under the average payment \eqref{eq:multi-all-to-all} is strictly
	IC and END for every $n\ge2$. Each respondent's honest ex-ante
	expected payment equals $C$. Every Bayesian equilibrium, including
	asymmetric and mixed equilibria with independent private
	randomizations, has finite interim expected payments and satisfies
	\begin{equation}\label{eq:multi-all-to-all-bound}
	\begin{aligned}
	 \sum_{i=1}^n\E[F_i^{(n)}]
	 &=nC-\frac1{n-1}\sum_{1\le i<\ell\le n}
	 \E\Bigl[\one_{\{J_i=J_\ell\}}
	 \bigl(w_{J_i}(Z_i)-w_{J_i}(Z_\ell)\bigr)\\[-2pt]
	 &\hspace{105pt}\cdot\bigl(G(Z_i)-G(Z_\ell)\bigr)\Bigr]
	 \le nC.
	\end{aligned}
	\end{equation}
	The identity \eqref{eq:multi-all-to-all-bound} holds at every equilibrium for any joint type law, whether exchangeable or not.
\end{theorem}
The proof is in Appendix~\ref{app:multi-end}.


\subsection{Infinitely many respondents}\label{sec:multi-infinity}

To compare to the original BTS we  consider now the case of infinitely many respondents. 
Let $(T_i)_{i\ge1}$ be an infinitely exchangeable sequence with common
pairwise distribution $P\in\mathcal E_m$. By de Finetti's theorem
there exists a  random variable  $\Theta\in\Delta_m$ conditional on which  the types are independent and identically distributed.
Suppose the respondents report simultaneously, using a common
(possibly mixed) reporting strategy conditional on type and independent
private randomizations. For each fixed respondent $i$ and $N>i$, define the following
statistics from the reports of the other $N-1$ respondents:
\begin{equation}\label{eq:multi-infty-finite-stats}
\begin{aligned}
 r_{i,k}^{(N)}
 &:=\frac{1}{N-1}\sum_{\substack{1\le\ell\le N\\\ell\ne i}}
       \one_{\{J_\ell=k\}},
 & H_i^{(N)}
 &:=\frac{1}{N-1}\sum_{\substack{1\le\ell\le N\\\ell\ne i}}
       G(Z_\ell),\\
 A_{i,j}^{(N)}
 &:=\frac{1}{N-1}\sum_{\substack{1\le\ell\le N\\\ell\ne i}}
       \one_{\{J_\ell=j\}}w_j(Z_\ell),
 & M_{i,j}^{(N)}
 &:=\frac{1}{N-1}\sum_{\substack{1\le\ell\le N\\\ell\ne i}}
       \one_{\{J_\ell=j\}}w_j(Z_\ell)Z_\ell,\\
 L_{i,j}^{(N)}
 &:=\frac{1}{N-1}\sum_{\substack{1\le\ell\le N\\\ell\ne i}}
       \one_{\{J_\ell=j\}}w_j(Z_\ell)G(Z_\ell).
\end{aligned}
\end{equation}
Let $r_i^{(N)}=(r_{i,1}^{(N)},\ldots,r_{i,m}^{(N)})$.
Here $r_{i,k}^{(N)}$ is the fraction of the other $N-1$ respondents
reporting answer  $k$, and $r_i^{(N)}$ is the distribution of
their reported answers. The quantity $H_i^{(N)}$ is the average value
of $G$ at their reported predictions. For each answer $j$,
$A_{i,j}^{(N)}$ is the average prediction-dependent weight contributed
by respondents reporting $j$; $M_{i,j}^{(N)}$ is the corresponding
weighted prediction vector, and $L_{i,j}^{(N)}$ is the corresponding
weighted average value of $G$. These three averages are over
\emph{all} other respondents, with zero contribution from those not
reporting $j$. The vector
$(r_i^{(N)}+M_{i,j}^{(N)})/(1+A_{i,j}^{(N)})$
combines the reported answer frequencies with these weighted
predictions and belongs to $\Delta_m$, since
$\sum_k M_{i,j}^{(N)}(k)=A_{i,j}^{(N)}$.

Conditional on $\Theta$, the other respondents' reports are i.i.d.;
hence the following limits exist almost surely:
\begin{equation}\label{eq:multi-infty-stats}
\begin{gathered}
 r_k:=\lim_{N\to\infty}r_{i,k}^{(N)},\qquad
 r:=(r_1,\ldots,r_m),\qquad
 H:=\lim_{N\to\infty}H_i^{(N)},\\
 A_j:=\lim_{N\to\infty}A_{i,j}^{(N)},\qquad
 M_j:=\lim_{N\to\infty}M_{i,j}^{(N)},\qquad
 L_j:=\lim_{N\to\infty}L_{i,j}^{(N)},\qquad
 \nu_j:=\frac{r+M_j}{1+A_j}\in\Delta_m.
\end{gathered}
\end{equation}
For the limiting game, a \emph{symmetric Bayesian equilibrium} is a
common type-dependent reporting strategy for which every type
maximizes its conditional expected payment $\E[F_i^\infty(j,z)\mid T_i=t]$.

\begin{proposition}[Infinite-population payment]\label{prop:multi-infty-payment}
Under the preceding assumptions, for every fixed respondent $i$ and
report $(j,z)$, the all-to-all payment \eqref{eq:multi-all-to-all}
converges almost surely as $N\to\infty$ to
\begin{equation}\label{eq:multi-infty-payment}
 \boxed{F_i^\infty(j,z)
 =C-H-L_j+(1+A_j)\bigl[G(\nu_j)-D_G(\nu_j,z)\bigr].}
\end{equation}
The corresponding mechanism remains strictly
IC, and honesty is END among all symmetric Bayesian equilibria of the
infinite-population game.

\end{proposition}
The proof is in Appendix~\ref{app:multi-infty-payment}.

Compared to Bayesian Truth Serum (Prelec, 2004) that uses the logarithmic score, our construction
uses a proper score with a matching-answer Bregman penalty and can
have bounded payments; its finite-population END bound applies for
every $n\ge2$ under the  assumptions
 of
Theorem~\ref{thm:multi-end}.

\section{IC mechanisms on reduced domains: (i) monotonicity under garbling; (ii) implementation of expected payoffs}\label{sec:garbling}
Throughout this section we assume $m=n=2$.
We first establish END for an information-sensitive
payment mechanism on $\Dp$, then show that its honest payment decreases
under independent common binary garbling. We then study implementation
of prescribed interim payments on a bounded subdomain of $\Dp$. 
 
 \subsection{END and decrease of honest payoffs under garbling}

We construct a mechanism that is END and strictly IC on $\Dp$.
We then consider how honest expected payments change when both respondents'
type signals are passed independently through the same stochastic channel,
in the sense of Blackwell (1953). Denote
\begin{equation}\label{eq:garb-parameters}
\begin{aligned}
\pi&:=\PP(T_i=1)=\frac{y}{1-x+y},\qquad
\rho:=x-y>0,\\
v&:=\operatorname{Cov}(T_1,T_2)=\pi(1-\pi)\rho>0.
\end{aligned}
\end{equation}
An \emph{independent common binary garbling} is a binary channel
$Q(s\mid t)=\PP(S_i=s\mid T_i=t)$, $s,t\in\{0,1\}$, satisfying
\[
\PP(S_1=s_1,S_2=s_2\mid T_1=t_1,T_2=t_2)
=Q(s_1\mid t_1)Q(s_2\mid t_2),\qquad
q:=\PP(S_i=1)\in\I.
\]
The case $Q(1\mid0)=Q(1\mid1)\in\I$ (independent outputs) is admitted;
$q\in\{0,1\}$ (constant outputs) and prediction reports
$z\in\{0,1\}$ are excluded.

Assume $G\in C([0,1])\cap C^2(\I)$ and $G''>0$ on $\I$.
Recall $S_G$ and $D_G$ from \eqref{eq:score} and \eqref{eq:bregman}.
For $\kappa,\lambda>0$, define 
\begin{equation}\label{eq:bonus}
 h(k,u):=2-\frac{k}{u}-\frac{1-k}{1-u},\qquad k\in\{0,1\},\ u\in\I,
\end{equation}
and the common payment functions
\begin{equation}\label{eq:mechanism}
 \boxed{\widetilde f_{jk}(z,u)=C+S_G(z,k)-G(u)
 -\kappa\one_{\{j=k\}}D_G(u,z)+\lambda h(k,u).}
\end{equation}

\begin{proposition}[Truthfulness and honest payments]\label{prop:honesty}
 The mechanism \eqref{eq:mechanism} is strictly IC on $\Ds$.
 Each respondent's honest ex-ante payment equals $C+\lambda M$, where, for $(x,y)\in\Ds$,
 \begin{equation}\label{eq:M}
 M:=2-\frac{\pi}{x}-\frac{1-\pi}{1-y}
 =\frac{v}{\pi^2+v}+\frac{v}{(1-\pi)^2+v}.
 \end{equation}
 Moreover, $M\in(0,1)$ on $\Dp$, whereas $M<0$ on $\Dm$.
\end{proposition}
The proof is in Appendix~\ref{app:garbling-results}.
The term $M$ depends on the joint distribution, but not on the generator $G$.


For any equilibrium, let $(J_i,Z_i)$ be respondent $i$'s report and
$\widetilde F_i:=\widetilde f_{J_iJ_{-i}}(Z_i,Z_{-i})$.

\begin{theorem}[Aggregate-payment bound]\label{thm:equilibrium}
 Every finite-payoff Bayesian equilibrium, including asymmetric and mixed
 equilibria, satisfies
 \begin{equation}\label{eq:mainbound}
 \boxed{\E[\widetilde F_1+\widetilde F_2]\le 2(C+\lambda M).}
 \end{equation}
 Equality holds only at honesty and at the strategy in which
 both respondents report the opposite answer and the corresponding
 probability of the peer reporting answer $1$, namely
 $(J_i,Z_i)=(1-T_i,1-p_{T_i})$, $i=1,2$.
 Thus, the mechanism is END on $\Dp$.
\end{theorem}
The proof is in Appendix~\ref{app:garbling-results}.

\noindent For $(x,y)\in\Dm$, $v<0$ and the denominators in
\eqref{eq:M} are positive, so $M<0$. The uninformative equilibrium
$U_r$ has expected payment $C$ under \eqref{eq:mechanism} because
$\sum_{k=0}^1r_kh(k,r)=0$ ($r_1=r$, $r_0=1-r$).
Thus the mechanism is strictly IC but not END anywhere on $\Dm$
when $\lambda>0$.

\begin{remark}
 A downside of this mechanism is that
 the bonus \eqref{eq:bonus} is unbounded below as $(k,u)\to(1,0)$ or
 $(0,1)$.
\end{remark}

\begin{theorem}[Decrease of honest payoffs under garbling]\label{thm:garbling}
 Fix $(x,y)\in\Dp$ and let $S_1,S_2\in\{0,1\}$ be obtained from
 $T_1,T_2$ by an independent common binary garbling $Q$ with
 \[
 \begin{gathered}
 \alpha:=Q(1\mid0),\qquad \delta:=Q(1\mid1)-Q(1\mid0),\\
 0\le\alpha\le1,\quad 0\le\alpha+\delta\le1,\quad
 q:=\alpha+\pi\delta\in(0,1).
 \end{gathered}
 \]
 Write
 \[
 \begin{aligned}
 q&=\PP(S_i=1)=\alpha+\pi\delta,\\
 c&:=\operatorname{Cov}(S_1,S_2)=\delta^2v,\\
 M_S&:=\frac{c}{q^2+c}+\frac{c}{(1-q)^2+c}.
 \end{aligned}
 \]
 Let $\widetilde F_i^{\,T}$ and $\widetilde F_i^{\,S}$ denote the payments
 under \eqref{eq:mechanism} when the respondents report honestly using,
 respectively, $T_i$ and $S_i$ and their corresponding conditional beliefs.
 Then, for $i=1,2$,
 \begin{equation}\label{eq:garbling-monotone-payoffs}
 \boxed{\E[\widetilde F_i^{\,S}]=C+\lambda M_S
 \le C+\lambda M=\E[\widetilde F_i^{\,T}].}
 \end{equation}
 Equality holds if and only if
 $(\alpha,\delta)=(0,1)$ (identity) or
 $(\alpha,\delta)=(1,-1)$ (the channel $S_i=1-T_i$).
 Otherwise the inequality is strict.
\end{theorem}
Thus, each respondent's honest ex-ante expected payment under the garbled type signals is no greater than under the original type signals.
The equality cases are the two deterministic bijections of the binary
signal, which preserve its information. If $\delta=0$, then
$S_1$ is independent of  $S_2$, $M_S=0$, and honest payment is $C$; honesty remains a
Bayesian equilibrium but is no longer strictly IC, since either answer is
optimal when the correct prediction $q$ is reported.

\paragraph{Intuition for garbling monotonicity.}
The two agreement probabilities are
\[
\PP(T_1=T_2=1)=\pi x=\pi^2+v,\qquad
\PP(T_1=T_2=0)=(1-\pi)(1-y)=(1-\pi)^2+v.
\]
Each fraction in \eqref{eq:M} measures the fraction of an agreement
probability attributable to covariance rather than independent
coincidence. The bonus depends only on the peer's report, so it does not alter a respondent's IC constraints. 
 As dependence vanishes, $M\to0$; as positive correlation tends
to one at a fixed interior marginal, $M\to1$.

\subsection{Implementation of non-constant expected payoffs}\label{Sect:impl}

This subsection concerns implementation of interim payments by  (discontinuous) IC mechanisms on the domain $\Dp^\epsilon$ defined below.
In some cases, the planner may want to implement expected payoffs dependent on $(x,y)$, for example those that reflect the information gain provided by the respondents as in the example below. Write the truthful interim payoff of type $t$ as
\[
V_t(x,y):=\E_{x,y}[F_i\mid T_i=t],\qquad t\in\{0,1\},
\]
using the same argument order $(x,y)$ for both types. For the results in this section, given $\epsilon\in(0,1/2)$, we restrict ourselves to
\[
\Dp^\epsilon:=\{(x,y)\in\Dp:\ \epsilon\le y<x\le 1-\epsilon\}.
\]
Similarly, the report space is
$\mathcal R_\epsilon=\{0,1\}\times I_\epsilon$, where
$I_\epsilon=[\epsilon,1-\epsilon]$. 
  
  For $V_1,V_0:\Dp^\epsilon\to\mathbb R$ define
  \begin{equation}\label{alphagamma}
  	\begin{aligned}
  		\alpha(x)&:=\sup_{\epsilon\le y<x\le 1-\epsilon}
  		\frac{(1-y)V_1(x,y)-(1-x)V_0(x,y)}{x-y},\\
  		\gamma(y)&:=\sup_{1-\epsilon\ge x>y\ge\epsilon}
  		\frac{x\,V_0(x,y)-y\,V_1(x,y)}{x-y}.
  	\end{aligned}
  \end{equation}
  Choose constants
  $a_1>\displaystyle\sup_{\epsilon<x\le1-\epsilon}\alpha(x)$ and
  $a_0>\displaystyle\sup_{\epsilon\le y<1-\epsilon}\gamma(y)$.
  On $I_\epsilon^2$ set
  \[
  \begin{aligned}
  	\widehat V_1(z,u)&=
  	\begin{cases}
  		V_1(z,u)-za_1-(1-z)a_0,&z>u,\\
  		0,&z\le u,
  	\end{cases}\\
  	\widehat V_0(z,u)&=
  	\begin{cases}
  		V_0(z,u)-ua_1-(1-u)a_0,&z>u,\\
  		0,&z\le u.
  	\end{cases}
  \end{aligned}
  \]
  Let $B=\max_{t=0,1}\sup_{I_\epsilon^2}|\widehat V_t|$ and choose
  \[
  L>\frac{B+B/\epsilon}{\epsilon}.
  \]
  For $z,u\in I_\epsilon$, define
  \begin{equation}\label{eq:implementation-mechanism}
  	\begin{aligned}
  		f_{11}(z,u)&=a_1-L\one_{\{z\ne u\}},&
  		f_{10}(z,u)&=a_0+\frac{\widehat V_1(z,u)}{1-z},\\
  		f_{01}(z,u)&=a_1+\frac{\widehat V_0(u,z)}{z},&
  		f_{00}(z,u)&=a_0-L\one_{\{z\ne u\}}.
  	\end{aligned}
  \end{equation}
We have the following result.
\begin{theorem}\label{P:pair}
 Let $V_1,V_0:\Dp^\epsilon\to\mathbb R$ be bounded.  If $\alpha$ and $\gamma$ are bounded above, the mechanism defined by \eqref{eq:implementation-mechanism}
 is strictly IC on $\Dp^\epsilon$ with report space $\mathcal R_\epsilon$,
and its truthful interim
  payoffs are $V_1(x,y)$ and $V_0(x,y)$.
\end{theorem}

The proof is in Appendix~\ref{app:implementation}. The implementing mechanism \eqref{eq:implementation-mechanism} is not $C^1$: the penalties $L\one_{\{z\ne u\}}$ are discontinuous along $z=u$.

\begin{proposition}[Necessary pointwise bounds for interim implementation]\label{prop:impl-necessary}
Suppose an IC mechanism on $\Dp^\epsilon$, with report space
$\mathcal R_\epsilon$, has finite honest interim payments $V_1,V_0$.
Then, we have
\[
\alpha(x)\le f_{11}(x,x)<\infty,
\qquad\gamma(y)\le f_{00}(y,y)<\infty.
\]
\end{proposition}
The proof is in Appendix~\ref{app:implementation}.
Whether pointwise finiteness (in place of the uniform boundedness
assumed in Theorem~\ref{P:pair}) suffices for implementation is open.

We now provide an example.

\begin{example}[Information gain]\label{ex:information-gain}\label{E:infogain}
 Suppose the planner wants the truthful interim payoff of a respondent of type
 $t\in\{0,1\}$ to equal the information gain
\[
V_t(x,y)=\sum_{k=0}^1\PP(T_{-i}=k\mid T_i=t)
\log\frac{\PP(T_{-i}=k\mid T_i=t)}{\PP(T_{-i}=k)},
\qquad p_1=x,\quad p_0=y,
\]
 where $\pi=\PP(T_i=1)=y/(1-x+y)$ as in \eqref{eq:garb-parameters}.
 It is straightforward to verify that
\[
\begin{aligned}
V_1(x,y)
&=x\log\frac{x}{\pi}
 +(1-x)\log\frac{1-x}{1-\pi}\\
&=x\log\frac{x}{y}+\log(1-x+y)>0,\\
V_0(x,y)
&=y\log\frac{y}{\pi}
 +(1-y)\log\frac{1-y}{1-\pi}\\
&=(1-y)\log\frac{1-y}{1-x}+\log(1-x+y)>0.
\end{aligned}
\]
On the compact triangle
$\{\epsilon\le y\le x\le1-\epsilon\}$,
$V_1,V_0$ are continuously differentiable and vanish on $x=y$.
The numerators in \eqref{alphagamma} are therefore Lipschitz
and vanish on the diagonal, so their ratios by $x-y$ are bounded
on $\Dp^\epsilon$. Thus the hypotheses of Theorem~\ref{P:pair} hold
for every $\epsilon\in(0,1/2)$.
\end{example}
\begin{remark}
		We restrict to $\Dp^\epsilon$ because the sufficient uniform bounds used in the construction need not hold for the information-gain example on the full domain $\Dp$. We do not know whether those interim payments can be implemented on the unrestricted $\Dp$ by another mechanism. 
		\end{remark}

\section{Conclusion}
For two respondents  answering a binary question  when the planner has no way to verify the answer,  if their types are exchangeable and   each respondent reports her answer and the probability she assigns to the other respondent's answer, we characterize all continuously differentiable payment rules that make honesty a best response to an honest peer.
In the  characterization, a proper scoring rule rewards the prediction, using the other respondent's answer to score it. 
The mechanism deducts a penalty when the two answers agree but the predictions differ. An analogous characterization is also obtained for two respondents and an arbitrary number of possible answers, under an additional regularity condition. The sufficient conditions for IC extend to an arbitrary number of respondents.

Making honesty a strict best response still leaves room for other equilibria.
 We therefore ask whether another equilibrium can make one respondent better off without making the other worse off, before either knows her signal. 
We find necessary conditions for the mechanisms that do not allow this, and construct a payment rule that rules out such an improvement. The construction extends to arbitrary finite numbers of respondents and answers. In the infinite-population limit, it remains strictly IC and satisfies END among symmetric Bayesian equilibria.

If types are positively correlated, the planner has more freedom. In the case $m=n=2$, we provide a mechanism under which each respondent earns less in expectation when the same random error is applied independently to both types and information is lost. No other  equilibrium has a higher total expected payment than honest reporting. 
Furthermore, a planner may want to specify what each respondent earns in expectation conditional on  her type. For example, the payment might reflect how much she learns about the other respondent's answer. We give conditions, in the positive correlation case,  under which the planner can implement these payments.


There are questions that remain unsolved.
  Characterizing $C^1$ IC mechanisms in the  $m\times n$ case with $n>2$  is an open problem.
A complete characterization of END mechanisms is also missing.
Whether one mechanism can deliver the prescribed payments and also protect honesty against more profitable equilibria remains open. It is also
unknown whether the information-gain interim payoffs of
Example~\ref{ex:information-gain} can be implemented on the full
$D_+$ domain without restricting beliefs and reports away from the boundary. 


\section*{References}
\begingroup\setlength{\parindent}{0pt}\setlength{\parskip}{0.5em}
\noindent Aznag, A., Bowers, R., Cummings, R., Hartline, J., vonAllmen, M. and Waggoner, B. (2026) Sample complexity of peer prediction. In \emph{Proceedings of the 27th ACM Conference on Economics and Computation (EC 2026)}. arXiv:2608.16838.\par
\noindent Baillon, A. (2017) Bayesian markets to elicit private information. \emph{Proceedings of the National Academy of Sciences} 114(30), 7958--7962.\par
\noindent Blackwell, D. (1953) Equivalent comparisons of experiments. \emph{Annals of Mathematical Statistics} 24(2), 265--272.\par
\noindent Cvitani\'c, J., Prelec, D., Radas, S. and \v{S}iki\'c, H. (2020) Incentive compatible surveys via posterior probabilities. \emph{Theory of Probability and Its Applications} 65(2), 292--321.\par
\noindent Cvitani\'c, J., Prelec, D., Riley, B. and Tereick, B. (2019) Honesty via choice-matching. \emph{American Economic Review: Insights} 1(2), 179--192.\par
\noindent Frongillo, R. and Kash, I. A. (2021) General truthfulness characterizations via convex analysis. \emph{Games and Economic Behavior} 130, 636--662.\par
\noindent Frongillo, R. and Witkowski, J. (2017) A geometric perspective on minimal peer prediction. \emph{ACM Transactions on Economics and Computation} 5(3), Article 17.\par
\noindent Han, Q., Schoenebeck, G., Tao, B. and Xia, L. (2025) Strong equilibria in Bayesian games with bounded group size. In \emph{Proceedings of the ACM Web Conference 2025}. doi:10.1145/3696410.3714585.\par
\noindent Kong, Y. (2024) Dominantly truthful peer prediction mechanisms with a finite number of tasks. \emph{Journal of the ACM} 71(2), Article 9. doi:10.1145/3638239.\par
\noindent Kong, Y. and Schoenebeck, G. (2018) Equilibrium selection in information elicitation without verification via information monotonicity. In \emph{Proceedings of the 9th Innovations in Theoretical Computer Science Conference (ITCS 2018)}, LIPIcs 94, 13:1--13:20.\par
\noindent Kong, Y. and Schoenebeck, G. (2019) An information theoretic framework for designing information elicitation mechanisms that reward truth-telling. \emph{ACM Transactions on Economics and Computation} 7(1), 2:1--2:33.\par
\noindent Lehmann, N. V. (2026) Mechanisms for belief elicitation without ground truth. \emph{Journal of Economic Surveys} 40(1), 505--527. doi:10.1111/joes.70000.\par
\noindent Peker, C. (2025) Optimal linear aggregation of correlated expert judgments. Available at SSRN 5385227. doi:10.2139/ssrn.5385227.\par
\noindent Prelec, D. (2004) A Bayesian truth serum for subjective data. \emph{Science} 306(5695), 462--466.\par
\noindent Prelec, D. (2021) Bilateral Bayesian truth serum: the $n\times m$ signals case. Available at SSRN 3908446.\par
\noindent Radanovic, G. and Faltings, B. (2013) A robust Bayesian truth serum for non-binary signals. In \emph{Proceedings of the 27th AAAI Conference on Artificial Intelligence (AAAI'13)}, 833--839.\par
\noindent Radanovic, G. and Faltings, B. (2014) Incentives for truthful information elicitation of continuous signals. In \emph{Proceedings of the 28th AAAI Conference on Artificial Intelligence (AAAI'14)}, 770--776.\par
\noindent Schoenebeck, G. and Yu, F.-Y. (2023) Two strongly truthful mechanisms for three heterogeneous agents answering one question. \emph{ACM Transactions on Economics and Computation} 10(4), Article 14.\par
\noindent Shnayder, V., Agarwal, A., Frongillo, R. and Parkes, D. C. (2016) Informed truthfulness in multi-task peer prediction. In \emph{Proceedings of the 17th ACM Conference on Economics and Computation (EC'16)}, 179--196.\par
\noindent Witkowski, J. and Parkes, D. C. (2012) A robust Bayesian truth serum for small populations. In \emph{Proceedings of the 26th AAAI Conference on Artificial Intelligence (AAAI'12)}, 1492--1498.\par
\noindent Zhang, Y., Xu, S., Schoenebeck, G. and Pennock, D. M. (2025) Stochastically dominant peer prediction. In \emph{Advances in Neural Information Processing Systems 38 (NeurIPS 2025)}.\par
\noindent Zhao, Z., Chen, X. and Zhou, Y. (2025) It takes two: A peer-prediction solution for blockchain verifier's dilemma. Available at SSRN 5399132. doi:10.2139/ssrn.5399132.\par
\endgroup

	\appendix
	\section{Proofs}\label{app:proofs}
\subsection{Proof of Theorem~\ref{thm:main}}\label{app:thm31}
\begin{proof}
Fix a respondent and omit the superscript $i$.

\textbf{Necessity.}
The first-order conditions for truthful predictions are
\[
\begin{aligned}
 x\partial_1 f_{11}(x,x)+(1-x)\partial_1 f_{10}(x,y)&=0,\\
 y\partial_1 f_{01}(y,x)+(1-y)\partial_1 f_{00}(y,y)&=0,
\end{aligned}\qquad (x,y)\in\Ds.
\]
The right-hand sides of
\[
 \partial_1 f_{10}(x,y)=-\frac{x}{1-x}\partial_1 f_{11}(x,x),\qquad
 \partial_1 f_{01}(y,x)=-\frac{1-y}{y}\partial_1 f_{00}(y,y)
\]
depend only on the first argument. By continuity across $z=u$
and integration,
\begin{equation}\label{eq:binary-offdiag-short}
 f_{10}(z,u)=a(z)+B_0(u),\qquad
 f_{01}(z,u)=c(z)+B_1(u),
\end{equation}
with $a,c,B_0,B_1\in C^1(\I)$.
Subtract $B_k(u)$ from $f_{jk}(z,u)$, obtaining
$\widetilde f_{jk}$, and set
$r_1(u)=\widetilde f_{11}(u,u)$,
$r_0(u)=\widetilde f_{00}(u,u)$. Define
\[
 G_1(u)=u r_1(u)+(1-u)a(u),\qquad
 G_0(u)=(1-u)r_0(u)+u c(u).
\]
IC for the deviations $(0,y)$ by type $1$ and $(1,x)$ by type $0$
gives
\[
 G_1(x)\ge xc(y)+(1-x)r_0(y),\qquad
 G_0(y)\ge(1-y)a(x)+yr_1(x).
\]
Letting $(x,y)\to(u,u)$ from $\Ds$ yields
$G_1(u)=G_0(u)=:G(u)$. Hence
\[
 r_1(u)=\frac{G(u)-(1-u)a(u)}u,\qquad
 r_0(u)=\frac{G(u)-u c(u)}{1-u}.
\]
The preceding IC inequalities become
\begin{align*}
 (1-y)G(x)-(1-x)G(y)-(x-y)c(y)&\ge0,\\
 xG(y)-yG(x)-(x-y)a(x)&\ge0.
\end{align*}
Both vanish at $x=y=u$. Differentiating at equality, using
both sides of $x=y$, gives
\begin{equation}\label{eq:binary-score-short}
 c(u)=G(u)+(1-u)G'(u)=S_G(u,1),\qquad
 a(u)=G(u)-uG'(u)=S_G(u,0).
\end{equation}
Thus $G'=c-a\in C^1(\I)$, so $G\in C^2(\I)$.
Substituting \eqref{eq:binary-score-short} into the first
imitation inequality gives
\[
 G(x)\ge x c(y)+(1-x)a(y)
       =G(y)+(x-y)G'(y),
\]
so $G$ is convex, and $r_1=c$, $r_0=a$.

Define
\[
 K_1(z,u):=c(z)-\widetilde f_{11}(z,u),\qquad
 K_0(z,u):=a(z)-\widetilde f_{00}(z,u).
\]
Then $K_j\in C^1(\I^2)$ and $K_j(u,u)=0$.
IC for deviations that change the answer only gives
\[
 (1-x)K_0(x,y)\ge0,\qquad
 yK_1(y,x)\ge0 \qquad ((x,y)\in\Ds).
\]
By continuity, $K_j(z,u)\ge0$ for all $(z,u)\in\I^2$.
These identities establish \eqref{eq:normal-form}--\eqref{eq:penalty-conds}.

\textbf{Sufficiency and strictness.}
Write $D_G(p,z)=G(p)-G(z)-(p-z)G'(z)$. The proper-score
identity $\E_p[S_G(z,k)]=G(p)-D_G(p,z)$ yields the loss
for a type-$t$ respondent from reporting $(j,z)$ against honesty:
\begin{equation}\label{eq:binary-loss-short}
 \mathcal L_t(j,z)
 =D_G(p_t,z)+\Pr(T_{-i}=j\mid T_i=t)K_j(z,p_j)\ge0.
\end{equation}
This is also the $m=2$ instance of \eqref{eq:multi-loss}.
For necessity of strictness, given any distinct $p,q\in\I$ choose
$(x,y)=(p,q)\in\Ds$. Strict IC for type $1$ against
$(0,q)$ implies $D_G(p,q)>0$, and against $(0,p)$ implies
$(1-p)K_0(p,q)>0$. Interchanging the types gives
$K_1(p,q)>0$ for all $p\ne q$. For differentiable convex $G$,
$D_G(p,q)>0$ for every $p\ne q$ is equivalent to strict
convexity. Conversely, strict convexity and $K_j(z,u)>0$
when $z\ne u$ make \eqref{eq:binary-loss-short} strictly
positive for each $(j,z)\ne(t,p_t)$, since $p_0\ne p_1$.
\end{proof}

\subsection{Verification of the continuous example in Section~\ref{sec:normal}}\label{app:nonc1}
For \eqref{eq:continuous-example}, put $L(a,b)=\sum_{h\in\{1,2,4\}}|a^h-b^h|$ and $g(a)=(1-a^2)^2$. Then
\[
g(a)-g(b)=(b^2-a^2)(2-a^2-b^2),\qquad
|g(a)-g(b)|\le2|a^2-b^2|\le2L(a,b).
\]
Let $\Lambda_t(j,z)$ be the loss of type $t$ from reporting $(j,z)$ against honesty. Direct substitution into \eqref{eq:continuous-example} gives
\begin{align*}
\Lambda_1(1,z)&=4xL(z,x)+(1-x)y[g(x)-g(z)],\\
\Lambda_1(0,z)&=(1-x)\bigl\{(x-y)[g(y)-g(x)]
+x[g(z)-g(y)]+4L(z,y)\bigr\},\\
\Lambda_0(0,z)&=y(1-x)[g(z)-g(y)]+4(1-y)L(z,y),\\
\Lambda_0(1,z)&=y\bigl\{(y-x)[g(x)-g(y)]
+4L(z,x)-(1-y)[g(z)-g(x)]\bigr\}.
\end{align*}
For $z\ge x$ the two terms of $\Lambda_1(1,z)$ are nonnegative;
for $z<x$, their possibly negative part is strictly smaller in absolute
value than $4xL(z,x)$. Similarly $\Lambda_0(0,z)>0$ for $z\ne y$,
using $2-z^2-y^2\le2(1-y)(1+y)$ when $z>y$.
Finally, $(x-y)[g(y)-g(x)]>0$ whenever $x\ne y$, and
$|g(z)-g(u)|\le2L(z,u)$ implies that the other terms in
$\Lambda_1(0,z)$ and $\Lambda_0(1,z)$ cannot offset these
strictly positive terms. Hence $\Lambda_t(j,z)>0$ for all
$(j,z)\ne(t,p_t)$. Boundedness and continuity are immediate.

\subsection{Proof of Theorem~\ref{thm:END-normal}}\label{app:thm41}
\begin{proof}
\textbf{Step 1: uninformative equilibria.}
By Theorem~\ref{thm:main}, with common payments,
\[
 f_{jk}(z,u)=S_G(z,k)+B_k(u)-\one_{\{j=k\}}K_j(z,u),
 \qquad K_j(u,u)=0.
\]
 Consider the mixed strategy of  reporting type one with probability $r\in\I$, and otherwise answering zero. That is,
denoting by $\delta_{(j,r)}$  unit mass at report $(j,r)$, 
 set 
\[
 \mu_r=(1-r)\delta_{(0,r)}+r\delta_{(1,r)}.
\]
We verify directly that this is the equilibrium profile $U_r$
defined before Theorem~\ref{thm:END-normal}.
Set $R(r):=G(r)+(1-r)B_0(r)+rB_1(r)$. For each answer $j$ and prediction $z$, the expected payment
against the report distribution $\mu_r$ is
\[
\begin{aligned}
\E_{\mu_r}[f_{jK}(z,r)]
&=G(r)-D_G(r,z)+(1-r)B_0(r)+rB_1(r)-r_jK_j(z,r)\\
&=R(r)-D_G(r,z)-r_jK_j(z,r)\le R(r),
\qquad r_1=r,\quad r_0=1-r.
\end{aligned}
\]
Equality holds at $z=r$ for either answer. Therefore both answers
with prediction $r$ are best responses, and $U_r$ is an equilibrium
for every $(x,y)\in\Ds$, directly from \eqref{eq:normal-form}.

\textbf{Step 2: END forces a common constant payoff.}
Let $R(r)$ denote the common interim expected payment at the
uninformative equilibrium $U_r$, for either respondent and either type. Then
\begin{align}\label{eq:R}
 R(r)
 &= (1-r)S_G(r,0)+rS_G(r,1)
 +(1-r)B_0(r)+rB_1(r)\notag\\
 &=G(r)+(1-r)B_0(r)+rB_1(r),
\end{align}
where $K_j(r,r)=0$.
Write $V_t(x,y):=\E[f_{t,T_{-i}}(p_t,p_{T_{-i}})\mid T_i=t]$ for the honest interim payment (common to both respondents). If $R(r)>R(s)$, choose $(x_n,y_n)\in\Ds$
 converging to $(s,s)$. By continuity, $V_t(x_n,y_n)$ converges to
 $R(s)<R(r)$ for $t=0,1$.
$\displaystyle R(r)>V_t(x_n,y_n)\ (t=0,1)$ for $n$ large, contradicting END. Thus
\[
 \boxed{R(u)\equiv C\quad\text{for }u\in(0,1)}.
\]

\textbf{Step 3: normalize the peer terms and compute honest payoffs.}
Define
\begin{equation}\label{eq:Delta-definition}
 \Delta(u):=B_0(u)-B_1(u).
\end{equation}
Then $R\equiv C$ is equivalent to
\begin{equation}\label{eq:Bnorm}
 \boxed{\begin{aligned}
 B_0(u)&=C-G(u)+u\Delta(u),\\
 B_1(u)&=C-G(u)-(1-u)\Delta(u).
 \end{aligned}}
\end{equation}
Let
\[
 \mathcal L(u):=G(u)-u\Delta(u)=C-B_0(u),\qquad
 \mathcal M(u):=G(u)+(1-u)\Delta(u)=C-B_1(u).
\]
Under honesty, $j=k$ implies $z=u=p_j$ and hence $K_j(z,u)=0$. Moreover,
\[
 \E[S_G(p_t,T_{-i})\mid T_i=t]=G(p_t).
\]
\begin{align*}
 V_1(x,y)
 &=G(x)+xB_1(x)+(1-x)B_0(y)
 =C+(1-x)[\mathcal L(x)-\mathcal L(y)],\\
 V_0(x,y)
 &=G(y)+yB_1(x)+(1-y)B_0(y)
 =C+y[\mathcal M(y)-\mathcal M(x)].
\end{align*}
Recall the common marginal $\pi=\PP(T_i=1)$, and let
\begin{equation}\label{eq:beta}
 \pi=\frac{y}{1-x+y},\qquad \beta=\pi(1-x)=(1-\pi)y>0.
\end{equation}
Let $W(x,y)$ denote either respondent's honest ex-ante expected payment.
Since $\mathcal M-\mathcal L=\Delta$,
\begin{align}\label{eq:W}
 W(x,y)
 &=\pi V_1(x,y)+(1-\pi)V_0(x,y)\notag\\
 &=C+\beta\bigl[\mathcal L(x)-\mathcal L(y)
 +\mathcal M(y)-\mathcal M(x)\bigr]\notag\\
 &=\boxed{C+\beta[\Delta(y)-\Delta(x)]}.
\end{align}

\textbf{Step 4: END on both signs forces $\Delta$ to be constant.}
Since $R\equiv C$, END implies
\[
 W(x,y)\ge C,
 \quad\text{which is equivalent to}\quad \Delta(y)\ge\Delta(x)
 \qquad\text{for }(x,y)\in\Ds.
\]
Applying the inequality to $(x,y),(y,x)\in\Ds$ gives
 $\Delta(y)\ge\Delta(x)$ and $\Delta(x)\ge\Delta(y)$. Hence
\[
 \boxed{\Delta(u)\equiv a_0\quad\text{for }u\in(0,1)}.
\]
By \eqref{eq:W}, $W(x,y)\equiv C$ on $\Ds$.

\textbf{Step 5: affine normalization.}
By \eqref{eq:Bnorm},
\[
 B_k(u)=C-G(u)+a_0(u-k),\qquad\text{for }k\in\{0,1\}.
\]
Set $\widehat G(u)=G(u)-a_0u$. Since
\[
 S_{\widehat G}(z,k)=S_G(z,k)-a_0k,
 \qquad -\widehat G(u)=-G(u)+a_0u,
 \qquad D_{\widehat G}=D_G,
\]
we obtain
\begin{align*}
 f_{jk}(z,u)
 &=S_G(z,k)+C-G(u)+a_0(u-k)-\one_{\{j=k\}}K_j(z,u)\\
 &=\boxed{C+S_{\widehat G}(z,k)-\widehat G(u)
 -\one_{\{j=k\}}K_j(z,u)},
\end{align*}
which is \eqref{eq:ENDnormal}; $\widehat G''=G''$, $D_{\widehat G}=D_G$.
\end{proof}

\subsection{Proof of Theorem~\ref{thm:multi-rep}}\label{app:multi-rep}
\begin{proof}
For $m=2$, the result follows from Theorem~\ref{thm:main} by the binary-to-simplex identification in Section~\ref{sec:multi-signals}. For the remainder of this proof, assume $m\ge3$.
Fix $i\in\{1,2\}$ and omit the superscript $i$. Recall
$\Delta_m^\circ$, $S_m$, $P_{t\cdot}$, $\pi_t(P)$, and $p_t(P)$
from \eqref{eq:multi-simplex}--\eqref{eq:multi-rows}. The IC condition
is \eqref{eq:multi-ic}.

\textbf{Step 1: reward function identities:} we want to show that, for $u,w\in\Delta_m^\circ$:
\begin{align*}
 f_{ak}(w,u)&=f_{bk}(w,u),
 &\quad& a,b,k\text{ pairwise distinct},\quad \frac{u_a}{u_b}=\frac{w_a}{w_b},\\
 f_{jk}(u,u)&=f_{kk}(u,u),
 && j\ne k.
\end{align*}
Let
\begin{equation}\label{eq:multi-A}
 \mathcal A:=\{A\in\mathbb R^{m\times m}:
 A=A^{\mathsf T},\ A_{st}>0\text{ for all }s,t\},\qquad
 r_t(A):=\sum_{\ell=1}^m A_{t\ell},\qquad
 p_t(A):=\frac{A_{t\cdot}}{r_t(A)}.
\end{equation}
Since $p_t(cA)=p_t(A)$ for $c>0$, density and continuity imply
that IC extends from $\mathcal E_m$ to every $A\in\mathcal A$.
Fix distinct $a,b\in\{1,\ldots,m\}$ and
$A^0\in\mathcal A$ satisfying $p_a(A^0)=p_b(A^0)=w$.
Such matrices exist: for every $w\in\Delta_m^\circ$,
$A^0=ww^{\mathsf T}\in\mathcal A$ gives
$p_t(A^0)=w$ for all $t$.
Let $u_\ell=p_\ell(A^0)$ and define
\begin{equation}\label{eq:multi-d}
 d_\ell:=f_{a\ell}(w,u_\ell)-f_{b\ell}(w,u_\ell).
\end{equation}
By IC for types $a$ and $b$, respectively,
\begin{equation}\label{eq:multi-d-ic}
 \sum_{\ell=1}^m w_\ell d_\ell\ge0,\qquad
 \sum_{\ell=1}^m w_\ell d_\ell\le0.
\end{equation}
Hence
\begin{equation}\label{eq:multi-d-zero}
 \sum_{\ell=1}^m w_\ell d_\ell=0.
\end{equation}
For $A\in\mathcal A$, abbreviate $p_c=p_c(A)$. IC for type
$a$ against the deviation $(b,p_b)$ and IC for type $b$ against
$(a,p_a)$ give, respectively,
\[
\begin{aligned}
&\sum_{\ell=1}^m p_{a,\ell}
  \bigl[f_{a\ell}(p_a,p_\ell)-f_{b\ell}(p_b,p_\ell)\bigr]\ge0,\\
&\sum_{\ell=1}^m p_{b,\ell}
  \bigl[f_{b\ell}(p_b,p_\ell)-f_{a\ell}(p_a,p_\ell)\bigr]\ge0.
\end{aligned}
\]
Adding these inequalities yields
\begin{equation}\label{eq:multi-Phi}
 \Phi(A):=\sum_{\ell=1}^m
 \bigl[p_a(A)_\ell-p_b(A)_\ell\bigr]
 \bigl[f_{a\ell}(p_a(A),p_\ell(A))
       -f_{b\ell}(p_b(A),p_\ell(A))\bigr]\ge0.
\end{equation}
Since $p_a(A^0)=p_b(A^0)=w$, we have $\Phi(A^0)=0$.

Let $H=H^{\mathsf T}$ and $A^\eta=A^0+\eta H$.
Since $A^0_{st}>0$, we have $A^\eta\in\mathcal A$ for all
sufficiently small $|\eta|$ (of either sign). Define
\begin{equation}\label{eq:multi-v}
 v_c:=\left.\frac{d}{d\eta}p_c(A^\eta)\right|_{\eta=0}
 =\frac{H_{c\cdot}-w\sum_\ell H_{c\ell}}{r_c(A^0)}
 \in S_m,\qquad c=a,b.
\end{equation}
The expression for $v_c$ follows from the quotient rule and
$A^0_{c\cdot}=r_c(A^0)w$:
\[
\begin{aligned}
 p_c(A^\eta)
 &=\frac{A^0_{c\cdot}+\eta H_{c\cdot}}
         {r_c(A^0)+\eta\sum_{\ell=1}^m H_{c\ell}},\\
 \left.\frac{d}{d\eta}p_c(A^\eta)\right|_{\eta=0}
 &=\frac{r_c(A^0)H_{c\cdot}
       -A^0_{c\cdot}\sum_{\ell=1}^m H_{c\ell}}
       {r_c(A^0)^2}.
\end{aligned}
\]
Moreover, $\sum_\ell(v_c)_\ell=0$, since
$\sum_\ell H_{c\ell}-\sum_\ell w_\ell\sum_q H_{cq}=0$.

By \eqref{eq:multi-d-zero}--\eqref{eq:multi-v},
\[
 \begin{aligned}
 p_a(A^\eta)-p_b(A^\eta)
    &=\eta(v_a-v_b)+o(\eta),\\
 f_{a\ell}(p_a(A^\eta),p_\ell(A^\eta))
 -f_{b\ell}(p_b(A^\eta),p_\ell(A^\eta))
    &=d_\ell+o(1).
 \end{aligned}
\]
Thus
\begin{equation}\label{eq:multi-pert}
 0\le\Phi(A^\eta)=\eta(v_a-v_b)\cdot d+o(\eta)
 \quad\text{for }\eta>0\text{ and }\eta<0.
\end{equation}
Since \eqref{eq:multi-pert} holds for both signs of $\eta$,
its first-order coefficient must vanish. Hence,
\[
(v_a-v_b)\cdot d=0.
\]
Let $E_{s\ell}$ be the matrix with a $1$ in position
$(s,\ell)$ and zeros elsewhere, and write $r_c=r_c(A^0)$.
For three symmetric perturbations, \eqref{eq:multi-v} gives
\[
\begin{array}{c|cc}
H &v_a&v_b\\ \hline
E_{aa} & (e_a-w)/r_a &0\\[2pt]
E_{a\ell}+E_{\ell a},\ \ell\notin\{a,b\}
   &(e_\ell-w)/r_a&0\\[2pt]
E_{ab}+E_{ba}&(e_b-w)/r_a&(e_a-w)/r_b
\end{array}
\]
where $e_s$ is the $s$th coordinate vector. Substituting the equality $(v_a-v_b)\cdot d=0$
obtained after \eqref{eq:multi-pert} yields
\begin{align}\label{eq:multi-pert-choices}
 H=E_{aa}\quad&\Longrightarrow\quad
     0=r_a^{-1}(d_a-w\cdot d),\nonumber\\
 H=E_{a\ell}+E_{\ell a},\ \ell\notin\{a,b\}
    \quad&\Longrightarrow\quad
     0=r_a^{-1}(d_\ell-w\cdot d),\\
 H=E_{ab}+E_{ba}\quad&\Longrightarrow\quad
     0=r_a^{-1}(d_b-w\cdot d)-r_b^{-1}(d_a-w\cdot d).
 \nonumber
\end{align}
Therefore, by \eqref{eq:multi-d-zero},
\begin{equation}\label{eq:multi-coincident}
 \boxed{f_{a\ell}(w,p_\ell(A^0))
       =f_{b\ell}(w,p_\ell(A^0)),\qquad \ell=1,\ldots,m.}
\end{equation}
For $a,b,k$ distinct, take $w,u\in\Delta_m^\circ$ with
$u_a/u_b=w_a/w_b$, and fix $c>0$. Prescribe the three rows
\begin{equation}\label{eq:multi-coincident-construction}
 A_{a\ell}=c w_a w_\ell,\qquad
 A_{b\ell}=c w_b w_\ell,\qquad
 A_{k\ell}=\frac{c w_a w_k}{u_a}u_\ell,
 \qquad \ell=1,\ldots,m.
\end{equation}
To obtain a symmetric matrix, entries prescribed more than once
must agree. There are exactly three such off-diagonal pairs:
\[
\begin{aligned}
 A_{ab}=c w_a w_b&=c w_b w_a=A_{ba},\\
 A_{ak}=c w_a w_k&=\frac{c w_a w_k}{u_a}u_a=A_{ka},\\
 A_{bk}=c w_b w_k&=\frac{c w_a w_k}{u_a}u_b=A_{kb},
 \qquad\left(\frac{u_b}{u_a}=\frac{w_b}{w_a}\right).
\end{aligned}
\]
Thus the prescriptions are consistent with symmetry. Explicitly,
for all remaining pairs $s,t\notin\{a,b,k\}$, set
\[
 A_{st}=A_{ts}:=c w_s w_t.
\]
Together with \eqref{eq:multi-coincident-construction} and
$A_{\ell s}=A_{s\ell}$ for $s\in\{a,b,k\}$, these formulas
specify every entry of $A$, all strictly positive and symmetric;
thus $A\in\mathcal A$. Its three prescribed row sums are
\[
 r_a(A)=c w_a,\qquad
 r_b(A)=c w_b,\qquad
 r_k(A)=\frac{c w_a w_k}{u_a},
\]
so $p_a(A)=p_b(A)=w$ and $p_k(A)=u$.
Equation~\eqref{eq:multi-coincident} therefore yields
\begin{equation}\label{eq:multi-merge}
 f_{ak}(w,u)=f_{bk}(w,u),\qquad
 a,b,k\text{ distinct},\quad u_a/u_b=w_a/w_b.
\end{equation}
Likewise, take $p_j=p_k=u$ in \eqref{eq:multi-coincident}; then
\begin{equation}\label{eq:multi-diagonal}
 f_{jk}(u,u)=f_{kk}(u,u),\qquad j\ne k.
\end{equation}

\textbf{Step 2: separability.}
Fix $j\ne k$, $z\in\Delta_m^\circ$, and
$r\in\mathbb R^m$ satisfying
\begin{equation}\label{eq:multi-radial}
 r_k=0,\quad r_\ell>0\ (\ell\ne k),\quad
 \sum_{\ell\ne k}r_\ell=1,\qquad
 u_\tau=(1-\tau)e_k+\tau r,\quad 0<\tau<1.
\end{equation}
Choose $A(\tau)\in\mathcal A$ symmetrically with
\begin{equation}\label{eq:multi-radial-A}
 A_{j\ell}=z_\ell\quad(1\le\ell\le m),\qquad
 A_{k\ell}=\frac{z_k}{r_j}r_\ell\quad(\ell\ne k),\qquad
 A_{kk}=\frac{z_k}{r_j}\frac{1-\tau}{\tau},
\end{equation}
and all other entries fixed and positive. Thus
\begin{equation}\label{eq:multi-radial-rows}
 p_j(A(\tau))=z,\qquad
 p_k(A(\tau))=u_\tau,\qquad
 p_\ell(A(\tau))\text{ is independent of }\tau\ \ (\ell\ne k).
\end{equation}
By IC, for each $\tau\in(0,1)$, the truthful prediction
$z=p_j(A(\tau))$ maximizes type $j$'s expected payment when
her reported answer is fixed at $j$:
\[
 z\in\operatorname*{arg\,max}_{\zeta\in\Delta_m^\circ}
 \sum_{\ell=1}^m z_\ell
 f_{j\ell}(\zeta,p_\ell(A(\tau))).
\]
Since $z\in\Delta_m^\circ$ and the payment functions are $C^1$,
the first-order condition with respect to the reported prediction,
for every $v\in S_m$, is
\begin{equation}\label{eq:multi-prediction-foc}
 0=\sum_{\ell=1}^m z_\ell
 \mathrm D_1 f_{j\ell}(z,p_\ell(A(\tau)))[v].
\end{equation}
By \eqref{eq:multi-radial-rows}, $z_\ell$ and
$p_\ell(A(\tau))$ for $\ell\ne k$ do not depend on $\tau$.
Thus the sum over $\ell\ne k$ in
\eqref{eq:multi-prediction-foc} is independent of $\tau$.
Since $z_k>0$, we conclude that, for all $\tau,\tau'\in(0,1)$,
\[
 \mathrm D_1 f_{jk}(z,u_\tau)[v]
 =\mathrm D_1 f_{jk}(z,u_{\tau'})[v].
\]
For $z,z_0\in\Delta_m^\circ$, integration on
$z_s=z_0+s(z-z_0)$, $0\le s\le1$, gives
\begin{equation}\label{eq:multi-radial-difference}
 \begin{aligned}
 &f_{jk}(z,u_\tau)-f_{jk}(z_0,u_\tau)\\
 &\quad=\int_0^1\mathrm D_1 f_{jk}(z_s,u_\tau)[z-z_0] \,ds
 =f_{jk}(z,u_{\tau'})-f_{jk}(z_0,u_{\tau'}).
 \end{aligned}
\end{equation}
Every $u\in\Delta_m^\circ$ is uniquely of the form
$u=(1-\tau)e_k+\tau r$, with $\tau=1-u_k$ and
$r_k=0$, $r_\ell=u_\ell/(1-u_k)$ for $\ell\ne k$.
By \eqref{eq:multi-vertex} with $a=b=j$ and
$u_\tau\to e_k$ as $\tau\downarrow0$, the difference
in \eqref{eq:multi-radial-difference} is independent of $u$.
Choose fixed $z_0,u_0\in\Delta_m^\circ$ and let
\begin{equation}\label{eq:multi-separation}
 s_{jk}(z):=f_{jk}(z,u_0)-f_{jk}(z_0,u_0),\qquad
 b_{jk}(u):=f_{jk}(z_0,u).
\end{equation}
Consequently,
\begin{equation}\label{eq:multi-sep}
 f_{jk}(z,u)=s_{jk}(z)+b_{jk}(u),\qquad j\ne k.
\end{equation}
For $a,b\ne k$, $a\ne b$, set
$u_h=(1-h)e_k+hw$, $w\in\Delta_m^\circ$. Then
$u_{h,a}/u_{h,b}=w_a/w_b$, so \eqref{eq:multi-merge} and
\eqref{eq:multi-vertex} imply
\begin{align}\label{eq:multi-combine}
 s_{ak}(w)-s_{bk}(w)
 &=b_{bk}(u_h)-b_{ak}(u_h)\nonumber\\
 &=-L_{ab,k},\qquad
 L_{ab,k}:=\lim_{h\downarrow0}
 [b_{ak}(u_h)-b_{bk}(u_h)]\in\mathbb R.
\end{align}
The limit $L_{ab,k}$ is independent of $w$, by
\eqref{eq:multi-vertex} and \eqref{eq:multi-separation}.
Taking $w=u$ in \eqref{eq:multi-merge} yields
\begin{equation}\label{eq:multi-coalesce}
 b_{ak}(u)-b_{bk}(u)=L_{ab,k}.
\end{equation}
Together with \eqref{eq:multi-combine} and \eqref{eq:multi-sep},
this implies
\[
 f_{ak}(z,u)=f_{bk}(z,u),\qquad a,b\ne k.
\]
Hence there are $s_k\in C^1(\Delta_m^\circ)$ and
$B_k\in C^1(\Delta_m^\circ)$ such that
\begin{equation}\label{eq:multi-offdiag}
 \boxed{f_{jk}(z,u)=s_k(z)+B_k(u),\qquad j\ne k.}
\end{equation}

\textbf{Step 3: score and penalties.}
Define
\begin{equation}\label{eq:multi-penalty-define}
 K_j(z,u):=s_j(z)+B_j(u)-f_{jj}(z,u).
\end{equation}
By \eqref{eq:multi-diagonal} and \eqref{eq:multi-offdiag},
we have $K_j(u,u)=0$.
For $q,w\in\Delta_m^\circ$, $q\ne w$, and $t\ne j$,
let $A\in\mathcal A$ satisfy
\begin{equation}\label{eq:multi-qw}
 A_{t\ell}=q_\ell,\qquad
 A_{j\ell}=\frac{q_j}{w_t}w_\ell,\qquad \ell=1,\ldots,m.
\end{equation}
The two prescriptions agree at the shared entry,
$A_{tj}=q_j=A_{jt}$. For $s,\ell\notin\{t,j\}$,
$s\ne\ell$, choose $A_{s\ell}=A_{\ell s}>0$ arbitrarily,
and choose $A_{ss}>0$ for each $s\notin\{t,j\}$.
Then $A=A^{\mathsf T}$, $A_{s\ell}>0$, and
\[
 p_t(A)=q,\qquad p_j(A)=w.
\]
For the following necessity argument, no further restriction
on these diagonal entries is needed: IC already holds on
$\mathcal A$ by Step~1. To obtain $A/\sum_{s,\ell}A_{s\ell}
\in\mathcal E_m$, as needed for the necessity of the
\emph{strict} IC conditions in Step~4, choose the free
$A_{ss}$ large enough that each remaining conditional row is
close to $e_s$; then all rows are pairwise distinct.
When $m=3$, there is only one free entry $A_{ss}$,
$s\notin\{t,j\}$. The map $A_{ss}\mapsto p_s(A)$ is
injective, so at most two values of $A_{ss}>0$ must be
excluded to ensure $p_s(A)\notin\{q,w\}$.
The deviation $(j,w)$ and
\eqref{eq:multi-offdiag}--\eqref{eq:multi-penalty-define} give
\begin{equation}\label{eq:multi-proper}
 q\cdot s(q)\ge q\cdot s(w),\qquad
 s(z):=(s_1(z),\ldots,s_m(z)).
\end{equation}
Define
\begin{equation}\label{eq:multi-G}
 G(q):=q\cdot s(q)=\sup_{w\in\Delta_m^\circ}q\cdot s(w).
\end{equation}
Thus $G$ is convex. For any $v\in S_m$, the first-order
condition for the maximization in \eqref{eq:multi-proper}
at $w=q$ gives
\begin{equation}\label{eq:multi-Gdiff}
 q\cdot\mathrm D s(q)[v]=0.
\end{equation}
By the product rule and \eqref{eq:multi-G},
\[
\mathrm D G(q)[v]
=v\cdot s(q)+q\cdot\mathrm D s(q)[v]=v\cdot s(q).
\]
Consequently, for each $k$,
\[
 S_G(q,k)=G(q)+\mathrm D G(q)[e_k-q]=s_k(q).
\]
Since $s_k\in C^1$, $G\in C^2(\Delta_m^\circ)$.
For the deviation $(j,q)$ by type $t\ne j$,
\begin{equation}\label{eq:multi-penalty-nonnegative}
 0\le U_t(t,q)-U_t(j,q)=q_jK_j(q,w).
\end{equation}
Since $q_j>0$, it follows that $K_j(q,w)\ge0$.
Continuity and $K_j(u,u)=0$ complete necessity.

\textbf{Step 4: sufficiency and strictness.}
Let
\begin{equation}\label{eq:multi-bregman}
 D_G(p,z):=G(p)-G(z)-\mathrm D G(z)[p-z]\ge0.
\end{equation}
For an honest peer, the type-$t$ loss from a report $(j,z)$ is
\begin{equation}\label{eq:multi-loss}
 U_t(t,p_t)-U_t(j,z)
 =D_G(p_t,z)+p_t(j)K_j(z,p_j)\ge0.
\end{equation}
For $n=2$, this proves IC sufficiency and, under the strict hypotheses, strict IC.
Conversely, \eqref{eq:multi-normal} implies
\eqref{eq:multi-vertex} because, for $a,b\ne k$,
\begin{equation}\label{eq:multi-vertex-converse}
 f_{ak}(z,u)-f_{bk}(z',u)=S_G(z,k)-S_G(z',k).
\end{equation}
For any $q\ne w$, take the pairwise-distinct-row completion
of \eqref{eq:multi-qw} described above, so that
$p_t=q$, $p_j=w$, $t\ne j$, and $P=A/\sum_{s,\ell}A_{s\ell}\in\mathcal E_m$.
Applying strict IC and \eqref{eq:multi-loss} to the deviations
$(j,w)$ and $(j,q)$, respectively, gives
\begin{align}\label{eq:multi-strict}
 D_G(q,w)&=U_t(t,q)-U_t(j,w)>0,\\
 q_jK_j(q,w)&=U_t(t,q)-U_t(j,q)>0.\nonumber
\end{align}
These conditions are equivalent to strict convexity of $G$
and $K_j(z,u)>0$ for $z\ne u$. Conversely, they make
\eqref{eq:multi-loss} strict for every $(j,z)\ne(t,p_t)$,
because $p_t\ne p_j$ for $t\ne j$ on $\mathcal E_m$.
\end{proof}

\subsection{Proof of Theorem~\ref{thm:multi-end}}\label{app:multi-end}
\begin{proof}
First, for the general-$n$ sufficiency of \eqref{eq:multi-normal}, conditional on $T_i=t$ each honest opponent has answer distribution $p_t$ and reports prediction $p_{T_\ell}$. Averaging over the $n-1$ opponents leaves the loss \eqref{eq:multi-loss} unchanged. This proves IC and, under the strict hypotheses, strict IC.

Let $\Delta_m=\{u\in[0,1]^m:\sum_k u_k=1\}$.
By compactness and the assumptions on $G$ and $\psi_j$,
\[
0<\underline{w}:=\min_{j,u\in\Delta_m}w_j(u)
\le w_j(u)\le \overline{w}:=\max_{j,u\in\Delta_m}w_j(u)<\infty.
\]
For $z\in\Delta_m^\circ$ and $v\in\Delta_m$, convexity
and continuity of $G$ give the supporting-plane inequality
\begin{equation}\label{eq:multi-end-boundary-bregman}
 D_G(v,z)=G(v)-G(z)-\mathrm DG(z)[v-z]\ge0.
\end{equation}
In fact $D_G(v,z)>0$ whenever $v\ne z$, even if
$v\in\partial\Delta_m$: otherwise equality in the
supporting-plane inequality at $z$ and $v$ forces $G$ to
be affine on $[z,v]$, hence on the nonempty open segment
$(z,v)\subset\Delta_m^\circ$, contradicting strict
convexity on $\Delta_m^\circ$.
Consequently,
\[
D_G(v,z)=0\ \Longleftrightarrow\ v=z,\qquad
K_j(z,u):=w_j(u)D_G(u,z)\ge0.
\]
Against a truthful peer, for type $t$ the loss from $(j,z)$ is
\[
D_G(p_t,z)+p_t(j)w_j(p_j)D_G(p_j,z)\ge0,
\]
with strict inequality for $(j,z)\ne(t,p_t)$, since $p_j\ne p_t$
when $j\ne t$. Thus \eqref{eq:multi-end-mechanism} is strictly IC.
The peer-only term $B_k(u)=C-G(u)$ is common to both respondents.
Moreover, for $a,b\ne k$,
\[
f_{ak}(z,u)-f_{bk}(z',u)=S_G(z,k)-S_G(z',k),
\]
so condition~\eqref{eq:multi-vertex} holds automatically.
This construction does not require condition~\eqref{eq:multi-vertex}
for either IC or END.
Under honesty, $J_i=T_i$, $Z_i=p_{T_i}(P)$, and
\begin{equation}\label{eq:multi-end-honest}
\begin{aligned}
\one_{\{T_i=T_{-i}\}}D_G(p_{T_{-i}},p_{T_i})&=0,\\
\E[S_G(p_{T_i},T_{-i})\mid T_i]&=G(p_{T_i}),\\
\E[F_i]&=C+\E[G(p_{T_i})]-\E[G(p_{T_{-i}})]=C,
\end{aligned}
\end{equation}
where the last equality follows from $P=P^{\mathsf T}$.

For $k\in\{1,\ldots,m\}$ and $z,u\in\Delta_m^\circ$,
let $y_s=(1-s)z+se_k\in\Delta_m^\circ$ for $0<s<1$.
The supporting-plane inequality and $G\in C(\Delta_m)$ yield
\[
G(y_s)\ge G(z)+\mathrm DG(z)[y_s-z]
\ \xrightarrow[s\uparrow1]{}\ S_G(z,k),\qquad
G(y_s)\xrightarrow[s\uparrow1]{}G(e_k).
\]
Hence
\begin{equation}\label{eq:multi-end-upper-bound}
 S_G(z,k)\le G(e_k),\qquad
 f_{jk}(z,u)\le C+\max_{\Delta_m}G-\min_{\Delta_m}G.
\end{equation}
For any fixed $z_0\in\Delta_m^\circ$,
$\inf_{j,k,u}f_{jk}(z_0,u)>-\infty$, since $G$ and $w_j$
are bounded on $\Delta_m$ and $D_G(\cdot,z_0)$ is continuous
on $\Delta_m$. Hence every equilibrium has finite interim payments.

Fix $i\in\{1,2\}$ and $T_i=t$; write
$(L,U)=(J_{-i},Z_{-i})$ for the peer's answer and prediction
under the conditional distribution given $T_i=t$.
A respondent's private randomization is independent of the
peer's type and private randomization conditional on $T_i$.
Thus, for every report $(j,z)$ used by type $t$,
\begin{equation}\label{eq:multi-end-independence}
 \mathcal L(L,U\mid T_i=t,J_i=j,Z_i=z)
 =\mathcal L(L,U\mid T_i=t).
\end{equation}
For every $j$ let
\begin{equation}\label{eq:multi-end-bestresponse}
\begin{aligned}
q&:=\E[e_L\mid T_i=t],\\
\alpha_j&:=\E[\one_{\{L=j\}}w_j(U)\mid T_i=t],\\
\mu_j&:=\E[\one_{\{L=j\}}w_j(U)U\mid T_i=t],\\
\nu_j&:=\frac{q+\mu_j}{1+\alpha_j}\in\Delta_m.
\end{aligned}
\end{equation}
The expected payoff from the report $(j,z)$ is
\begin{equation}\label{eq:multi-end-Phi}
\begin{aligned}
\mathcal U_{t,j}(z)
&=C-\E[G(U)\mid T_i=t]
-\E[\one_{\{L=j\}}w_j(U)G(U)\mid T_i=t]\\
&\qquad +(1+\alpha_j)\bigl[G(\nu_j)-D_G(\nu_j,z)\bigr].
\end{aligned}
\end{equation}
Since $D_G(\nu_j,z)>0$ for $\nu_j\ne z$, strict convexity implies
\[
\operatorname*{arg\,max}_{z\in\Delta_m^\circ}\mathcal U_{t,j}(z)
=\begin{cases}\{\nu_j\},&\nu_j\in\Delta_m^\circ,\\
\varnothing,&\nu_j\in\partial\Delta_m.
\end{cases}
\]
For the boundary case, fix $z_\star\in\Delta_m^\circ$ and
$z_s=\nu_j+s(z_\star-\nu_j)$; convexity gives, for $0<s<1/2$,
\[
0\le D_G(\nu_j,z_s)
\le G(\nu_j)-2G(z_s)+G(z_{2s})\longrightarrow0
\quad(s\downarrow0).
\]
Thus each equilibrium type uses at most one prediction per answer,
so $\operatorname{supp}(J_i,Z_i\mid T_i=t)$ has at most $m$ atoms.
The unconditional report support has at most $m^2$ atoms.
In particular $\nabla_{S_m}G(Z_i)$ is bounded there.

Define, with $\gamma_i\in S_m$ the gradient of $G$ on the simplex,
\begin{equation}\label{eq:multi-end-d}
\begin{aligned}
d_i&:=e_{J_{-i}}-Z_i
-\one_{\{J_i=J_{-i}\}}w_{J_i}(Z_{-i})(Z_i-Z_{-i}),\\
\gamma_i&:=\nabla_{S_m}G(Z_i).
\end{aligned}
\end{equation}
By \eqref{eq:multi-end-independence}, for every equilibrium
report $(J_i,Z_i)=(j,z)$ of type $t$,
\begin{equation}\label{eq:multi-end-FOC}
 \E[d_i\mid T_i=t,J_i=j,Z_i=z]
 =q+\mu_j-(1+\alpha_j)z=0.
\end{equation}
Since all reports have finite support in equilibrium,
$\gamma_i$ and $d_i$ are bounded on that support
(the weights $w_j$ are bounded on $\Delta_m$).
Consequently all products below are integrable and
\begin{equation}\label{eq:multi-end-orthogonality}
 \E\langle\gamma_i,d_i\rangle
 =\E\!\left\langle\gamma_i,
   \E[d_i\mid T_i,J_i,Z_i]\right\rangle=0.
\end{equation}
Let $I:=\one_{\{J_1=J_2\}}$ and write $j=J_1=J_2$ on $\{I=1\}$.
Using the Bregman identity and \eqref{eq:multi-end-d},
\begin{equation}\label{eq:multi-end-identity}
\begin{aligned}
F_1+F_2-2C
&=\langle\gamma_1,d_1\rangle+\langle\gamma_2,d_2\rangle\\
&\quad-I\bigl[w_j(Z_1)-w_j(Z_2)\bigr]
              \bigl[G(Z_1)-G(Z_2)\bigr].
\end{aligned}
\end{equation}
Taking expectations and using \eqref{eq:multi-end-orthogonality} gives
\[
\E[F_1^{(2)}+F_2^{(2)}]
=2C-\E\!\left[\one_{\{J_1=J_2\}}
\bigl(w_{J_1}(Z_1)-w_{J_1}(Z_2)\bigr)
\bigl(G(Z_1)-G(Z_2)\bigr)\right]\le2C,
\]
because
\[
\bigl[w_j(z)-w_j(u)\bigr]\bigl[G(z)-G(u)\bigr]
=\bigl[\psi_j(G(z))-\psi_j(G(u))\bigr]
 \bigl[G(z)-G(u)\bigr]\ge0.
\]
By \eqref{eq:multi-end-honest}, every honest payment is $C$.
Consequently no alternative equilibrium can weakly increase both
respondents' expected payments with one strict increase, and END follows
for $n=2$.

\medskip
\noindent\textbf{General $n$.}
Every opponent has conditional type distribution $p_t$ given $T_i=t$.
Thus, against honest opponents, the expected payoff difference between
$(t,p_t)$ and $(j,z)$ is given by the two-respondent calculation above:
\[
 D_G(p_t,z)+p_t(j)w_j(p_j)D_G(p_j,z),
\]
which is strictly positive for $(j,z)\ne(t,p_t)$. By exchangeability,
each respondent's honest expected payment equals $C$.

Fix an arbitrary Bayesian equilibrium. For each respondent $i$ and type
$t$, select $R_i$ uniformly from $\{1,\ldots,n\}\setminus\{i\}$,
independently of types and reporting randomizations. In the best-response
calculation, replace the peer's report $(L,U)$ in
\eqref{eq:multi-end-bestresponse}--\eqref{eq:multi-end-Phi} by
$(J_{R_i},Z_{R_i})$ conditional on $T_i=t$. The expected payoff from a
report $(j,z)$ is exactly \eqref{eq:multi-end-Phi}. Consequently, by the preceding two-respondent best-response argument, each answer used by a given type is associated
with a unique optimal prediction, and the equilibrium report support
of each respondent is finite. The boundedness argument above
shows that every equilibrium has finite interim expected payments.

For distinct $i,\ell$, set
\begin{equation}\label{eq:multi-n-d}
\begin{aligned}
 d_{i\ell}&:=e_{J_\ell}-Z_i
 -\one_{\{J_i=J_\ell\}}w_{J_i}(Z_\ell)(Z_i-Z_\ell),\\
 d_i&:=\frac1{n-1}\sum_{\ell\ne i}d_{i\ell},
 \qquad \gamma_i:=\nabla_{S_m}G(Z_i).
\end{aligned}
\end{equation}
Independence of private randomizations implies that the conditional
law of the other respondents' reports given $(T_i,J_i,Z_i)$ is their
law given $T_i$. The unique optimal-prediction condition in
\eqref{eq:multi-end-Phi}, applied to the uniformly selected opponent,
therefore yields
\begin{equation}\label{eq:multi-n-foc}
 \E[d_i\mid T_i,J_i,Z_i]=0,
 \qquad \E\langle\gamma_i,d_i\rangle=0.
\end{equation}
The last equality follows because finite equilibrium supports make
$\gamma_i$ bounded, while $d_i$ is bounded by boundedness of the weights.

For each $i<\ell$, the same Bregman identity as in
\eqref{eq:multi-end-identity} gives
\begin{equation}\label{eq:multi-n-pair}
\begin{aligned}
 &f_{J_iJ_\ell}(Z_i,Z_\ell)
 +f_{J_\ell J_i}(Z_\ell,Z_i)-2C\\
 &\quad=\langle\gamma_i,d_{i\ell}\rangle
 +\langle\gamma_\ell,d_{\ell i}\rangle
 -\one_{\{J_i=J_\ell\}}
 \bigl[w_{J_i}(Z_i)-w_{J_i}(Z_\ell)\bigr]
 \bigl[G(Z_i)-G(Z_\ell)\bigr].
\end{aligned}
\end{equation}
Summing \eqref{eq:multi-n-pair} over $i<\ell$, dividing by $n-1$,
and taking expectations gives \eqref{eq:multi-all-to-all-bound} by
\eqref{eq:multi-n-foc}, since
\[
 [w_j(z)-w_j(u)][G(z)-G(u)]
 =[\psi_j(G(z))-\psi_j(G(u))][G(z)-G(u)]\ge0.
\]
The bound and the honest payment $C$ establish END.
\end{proof}

\subsection{Proof of Proposition~\ref{prop:multi-infty-payment}}\label{app:multi-infty-payment}
\begin{proof}
Write $r_i^{(N)}=(r_{i,1}^{(N)},\ldots,r_{i,m}^{(N)})$ and
$\nu_{i,j}^{(N)}=(r_i^{(N)}+M_{i,j}^{(N)})/(1+A_{i,j}^{(N)})$,
with the finite-sample statistics defined in
\eqref{eq:multi-infty-finite-stats}.
Expanding $S_G$ and $D_G$ in \eqref{eq:multi-all-to-all} yields
\[
\begin{aligned}
 F_i^{(N)}(j,z)
 &=C-H_i^{(N)}-L_{i,j}^{(N)}+(1+A_{i,j}^{(N)})G(z)\\
 &\quad+\mathrm DG(z)\bigl[r_i^{(N)}+M_{i,j}^{(N)}
              -(1+A_{i,j}^{(N)})z\bigr]\\
 &=C-H_i^{(N)}-L_{i,j}^{(N)}
 +(1+A_{i,j}^{(N)})\bigl[G(\nu_{i,j}^{(N)})
                       -D_G(\nu_{i,j}^{(N)},z)\bigr].
\end{aligned}
\]
Given $\Theta$, the other respondents' reports are i.i.d.
The conditional strong law, boundedness of $G$ and $w_j$, and
continuity of $D_G(\cdot,z)$ on $\Delta_m$ imply that the
finite-sample statistics converge as in \eqref{eq:multi-infty-stats}.
Taking limits in the preceding identity gives
\eqref{eq:multi-infty-payment}. 

To verify the equilibrium claim,
fix any common reporting strategy and type $t$.
For each $N\ge2$, exchangeability and the independence of
private randomizations give, for any fixed report $(j,z)$,
\[
\E[F_i^{(N)}(j,z)\mid T_i=t]
=\E[f_{jJ_\ell}(z,Z_\ell)\mid T_i=t],\qquad \ell\ne i.
\]
The right-hand side is independent of $N$ and equals the interim
expected payment from the same report in the corresponding two-respondent
game. The limit-game best-response problem is thus identical to the
two-respondent problem. Hence strict IC persists, and every symmetric
limit equilibrium has expected payment at most $C$ by
\eqref{eq:multi-all-to-all-bound} for $n=2$.
\end{proof}

\subsection{Auxiliary lemmas for Section~\ref{sec:garbling}}\label{app:garbling-lemmas}

Throughout this subsection, fix $(x,y)\in\Dp$ and recall
$\pi=\PP(T_i=1)=y/(1-x+y)$,
$v=\operatorname{Cov}(T_1,T_2)=\pi(1-\pi)(x-y)>0$,
and $M$ from \eqref{eq:M}.

\begin{lemma}[Binary-channel contraction]\label{lem:channel}
 Let $J_i$ be a binary report based only on $T_i$, with independent private
 randomizations conditional on $(T_1,T_2)$. Set
 \[
 q_i:=\PP(J_i=1),\quad c:=\operatorname{Cov}(J_1,J_2),\quad
 A_j:=\PP(J_1=J_2=j)>0.
 \]
 Then
 \begin{equation}\label{eq:channelbound}
 \frac{c}{A_1}+\frac{c}{A_0}\le M.
 \end{equation}
 Equality holds if and only if $J_1=T_1$ and $J_2=T_2$ almost surely,
 or $J_1=1-T_1$ and $J_2=1-T_2$ almost surely.
\end{lemma}
\begin{proof}
 Let $\alpha_i:=\PP(J_i=1\mid T_i=0)$ and
 $\delta_i:=\PP(J_i=1\mid T_i=1)-\alpha_i$. Then
 \begin{equation}\label{eq:channelmoments}
 q_i=\alpha_i+\pi\delta_i,\quad
 c=\delta_1\delta_2v,\quad
 A_1=q_1q_2+c,\quad A_0=(1-q_1)(1-q_2)+c.
 \end{equation}
 Since $v>0$, if $c\le0$, then $c/A_1+c/A_0\le0<M$.
 For $c>0$, $\delta_1\delta_2>0$. If both $\delta_i<0$,
 replace each reported answer by $J_i':=1-J_i$, $i=1,2$. Then
 \[
 q_i'=1-q_i,\qquad \delta_i'=-\delta_i>0,\qquad
 A_1'=A_0,\quad A_0'=A_1,\quad c'=c.
 \]
 Thus $c/A_1+c/A_0$ is unchanged, and it suffices to prove
 the inequality when $\delta_1,\delta_2>0$.
 Since $0\le\alpha_i$ and $\alpha_i+\delta_i\le1$,
 \[
 q_i\ge\pi\delta_i,\qquad 1-q_i\ge(1-\pi)\delta_i.
 \]
 Consequently,
 \begin{align*}
 A_1&=q_1q_2+\delta_1\delta_2v
 \ge\delta_1\delta_2(\pi^2+v),\\
 A_0&=(1-q_1)(1-q_2)+\delta_1\delta_2v
 \ge\delta_1\delta_2((1-\pi)^2+v),\\
 \frac c{A_1}+\frac c{A_0}
 &\le\frac v{\pi^2+v}+\frac v{(1-\pi)^2+v}=M.
 \end{align*}
 With $c>0$, equality in both denominator inequalities implies
 $q_i=\pi\delta_i$ and $1-q_i=(1-\pi)\delta_i$ for $i=1,2$,
 hence $\alpha_i=0$ and $\delta_i=1$.
 If the original $\delta_1,\delta_2$ were negative, the
 preceding argument applies to $J_i'=1-J_i$ and yields
 $J_i'=T_i$, so $J_i=1-T_i$ for both respondents.
 These are exactly the two equality cases.
\end{proof}

Let $f^0$ denote the base mechanism without the peer-only bonus:
\[
f^0_{jk}(z,u):=
\widetilde f_{jk}(z,u)-\lambda h(k,u).
\]

\begin{lemma}[Best responses and finite support]\label{lem:support}
 Every finite-payoff equilibrium of \eqref{eq:mechanism} uses at most one
 prediction for each answer, for each type. Its best-response sets coincide
 with those of $f^0$. For every report used in equilibrium,
 \begin{equation}\label{eq:FOC}
 \E[d_i\mid T_i,J_i,Z_i]=0,\qquad
 d_i:=J_{-i}-Z_i-\kappa\one_{\{J_i=J_{-i}\}}(Z_i-Z_{-i}).
 \end{equation}
 Conversely, any equilibrium of $f^0$ is a finite-support equilibrium of
 \eqref{eq:mechanism}.
\end{lemma}
\begin{proof}
 For $k\in\{0,1\}$ and $z,u\in\I$,
 \[
 S_G(z,k)\le G(k),\qquad h(k,u)<1,\qquad D_G(u,z)\ge0.
 \]
 Since $G\in C([0,1])$, both $f^0$ and $\widetilde f$ are
 bounded above. Finite interim equilibrium payoffs therefore imply
 \[
 \E[f^0_{J_iJ_{-i}}(Z_i,Z_{-i})\mid T_i=t]\in\mathbb R,\qquad
 \E[h(J_{-i},Z_{-i})\mid T_i=t]\in\mathbb R.
 \]
 The second term is independent of respondent $i$'s report, hence the
 best-response sets coincide with those of $f^0$. As $f^0$ is \eqref{eq:family} with $w_j\equiv\kappa$,
the best-response calculation in Appendix~\ref{app:multi-end}
(proof of Theorem~\ref{thm:multi-end}, with $m=n=2$)
gives at most one optimal prediction per answer and finite
equilibrium support. At every used report $(j,z)$,
 with $(K,U)$ the peer's conditional report given $T_i=t$,
 \[
 \begin{aligned}
 0&=\PP(K=1\mid T_i=t)
 +\kappa\E[U\one_{\{K=j\}}\mid T_i=t]\\
 &\quad-\bigl(1+\kappa\PP(K=j\mid T_i=t)\bigr)z
 =\E[d_i\mid T_i=t,J_i=j,Z_i=z].
 \end{aligned}
 \]
 Conversely, every equilibrium of $f^0$ has finite interim payoffs and
 finite support by Appendix~\ref{app:multi-end}; on that support the
 peer-only bonus has finite conditional expectation, so the
 best responses remain unchanged.
\end{proof}

\begin{lemma}[Aggregate base payments and answer balance]\label{lem:budget}
 At every finite-payoff equilibrium of \eqref{eq:mechanism},
 \begin{equation}\label{eq:basebudget}
 \E[F_1^0+F_2^0]=2C,\qquad
 F_i^0:=f^0_{J_iJ_{-i}}(Z_i,Z_{-i}).
 \end{equation}
 Define $\chi_i:=J_iZ_i+(1-J_i)(1-Z_i)\in(0,1)$ and
 $A_j:=\PP(J_1=J_2=j)$. Then $A_0,A_1>0$ and
 \begin{equation}\label{eq:balance}
 \E[\one_{\{J_1=j\}}\chi_1+\one_{\{J_2=j\}}\chi_2]=2A_j,
 \qquad\text{for }j\in\{0,1\}.
 \end{equation}
\end{lemma}
\begin{proof}
 By Lemma~\ref{lem:support}, every finite-payoff equilibrium of
 $\widetilde f$ is an equilibrium of $f^0$, which is
 \eqref{eq:family} with $w_j\equiv\kappa$.
 Thus \eqref{eq:budget} implies \eqref{eq:basebudget}.
 On $\{J_i=j\}$, multiply \eqref{eq:FOC} by $2j-1$:
 \[
 \E\left[\one_{\{J_{-i}=j\}}-\chi_i
 -\kappa\one_{\{J_{-i}=j\}}(\chi_i-\chi_{-i})
 \mid T_i,J_i=j,Z_i\right]=0.
 \]
 Multiply by $\one_{\{J_i=j\}}$, average, and sum over $i=1,2$.
 The two matching terms cancel, giving \eqref{eq:balance}.
 If $A_j=0$, then \eqref{eq:balance} and $\chi_i>0$ imply
 $\PP(J_1=j)=\PP(J_2=j)=0$. Thus
 $\PP(J_1=1-j,J_2=1-j)=1$, so $A_{1-j}=1$ and
 \eqref{eq:balance} for $1-j$ implies $\E[\chi_1+\chi_2]=2$, contrary to
 $\chi_i<1$. Hence $A_0A_1>0$.
\end{proof}

\subsection{Proofs of Section~\ref{sec:garbling}}\label{app:garbling-results}

\begin{proof}[Proof of Proposition~\ref{prop:honesty}]
 Fix $(x,y)\in\Ds$. Recall $p_1=x$, $p_0=y$,
 $\pi=\PP(T_i=1)=y/(1-x+y)$, and
 $v=\pi(1-\pi)(x-y)$ (of either sign). Let
 $U_t(j,z):=\E[\widetilde f_{j,T_{-i}}(z,p_{T_{-i}})\mid T_i=t]$
 be the expected payment of type $t$ against an honest peer. Then
 \begin{align*}
 &U_t(t,p_t)-U_t(j,z)\\
 &\qquad=D_G(p_t,z)+\kappa\E\!\left[
 \one_{\{j=T_{-i}\}}D_G(p_{T_{-i}},z)\mid T_i=t\right]\ge0.
 \end{align*}
 If $z\ne p_t$, $D_G(p_t,z)>0$; if $z=p_t$ and $j\ne t$,
 \[
 U_1(1,x)-U_1(0,x)=\kappa(1-x)D_G(y,x)>0,\quad
 U_0(0,y)-U_0(1,y)=\kappa yD_G(x,y)>0.
 \]
 Thus honesty is strict. Its matching penalty is zero:
 \[
 \one_{\{T_i=T_{-i}\}}D_G(p_{T_{-i}},p_{T_i})=0.
 \]
 Since $\E[S_G(p_{T_i},T_{-i})\mid T_i]=G(p_{T_i})$ and
 $T_i\overset d=T_{-i}$,
 \[
 \E[F_i^0]=C+\E[G(p_{T_i})]-\E[G(p_{T_{-i}})]=C.
 \]
 Under honesty,
 \begin{align*}
 \E[h(T_{-i},p_{T_{-i}})]
 &=2-\frac{\PP(T_{-i}=1)}x
     -\frac{\PP(T_{-i}=0)}{1-y}
 =2-\frac\pi x-\frac{1-\pi}{1-y},\\
 \pi x&=\pi^2+v,\qquad
 (1-\pi)(1-y)=(1-\pi)^2+v,\\
 M&=\frac v{\pi^2+v}+\frac v{(1-\pi)^2+v}.
\end{align*}
For $(x,y)\in\Dp$, $0<v<\pi(1-\pi)$, so
\[
0<\frac{v}{\pi^2+v}<1-\pi,
\qquad 0<\frac{v}{(1-\pi)^2+v}<\pi,
\qquad 0<M<1.
\]

 Therefore $\E[\widetilde F_i]=C+\lambda M$.
\end{proof}

\begin{proof}[Proof of Theorem~\ref{thm:equilibrium}]
 Recall $M$ from \eqref{eq:M} and the reports $(J_i,Z_i)$ from
 Section~\ref{sec:model}. Let
 $q_i:=\PP(J_i=1)$, $c:=\operatorname{Cov}(J_1,J_2)$,
 $A_j:=\PP(J_1=J_2=j)$, and
 $\chi_i:=J_iZ_i+(1-J_i)(1-Z_i)$, as in Lemma~\ref{lem:budget}.
 For $j=0,1$, \eqref{eq:balance} and Cauchy--Schwarz yield
 \[
 \E\!\left[\frac{\one_{\{J_1=j\}}}{\chi_1}
 +\frac{\one_{\{J_2=j\}}}{\chi_2}\right]
 \ge\frac{[\PP(J_1=j)+\PP(J_2=j)]^2}{2A_j}.
 \]
 As $h(J_i,Z_i)=2-\chi_i^{-1}$,
 \begin{align}
 \frac12\E[h(J_1,Z_1)+h(J_2,Z_2)]
 &\le2-\frac{(q_1+q_2)^2}{4A_1}
        -\frac{(2-q_1-q_2)^2}{4A_0}\notag\\
 &\le2-\frac{q_1q_2}{A_1}
        -\frac{(1-q_1)(1-q_2)}{A_0}\notag\\
 &=\frac c{A_1}+\frac c{A_0}\le M.
 \label{eq:bonusbound}
 \end{align}
 By \eqref{eq:basebudget} and independence of the bonus from the respondent's
 report,
 \[
 \E[\widetilde F_1+\widetilde F_2]
 =2C+\lambda\E[h(J_1,Z_1)+h(J_2,Z_2)]\le2(C+\lambda M).
 \]
 Equality in \eqref{eq:bonusbound} implies, by Lemma~\ref{lem:channel},
 $J_i=T_i$ for both respondents or $J_i=1-T_i$ for both.
 In these equality cases, let $z_{it}$ be respondent $i$'s prediction when
 $T_i=t$, and write $q_t:=\PP(J_{-i}=1\mid T_i=t)$ and
 $r_t:=\PP(J_{-i}=J_i\mid T_i=t)$. The prediction equations are
 \[
 \begin{pmatrix}1+\kappa r_t&-\kappa r_t\\
 -\kappa r_t&1+\kappa r_t\end{pmatrix}
 \begin{pmatrix}z_{1t}\\z_{2t}\end{pmatrix}
 =\begin{pmatrix}q_t\\q_t\end{pmatrix},
 \qquad\det=1+2\kappa r_t>0.
 \]
 Hence $z_{1t}=z_{2t}=q_t$, yielding either honest reports
 $(J_i,Z_i)=(T_i,p_{T_i})$ for both respondents, or
 $(J_i,Z_i)=(1-T_i,1-p_{T_i})$ for both respondents.
 To verify that the second profile is an equilibrium, fix type $t$,
write $U_t(j,z)$ for the type-$t$ expected payment from $(j,z)$,
let $(K,U)$ denote the peer's report, and put
$r_j=\PP(K=j\mid T_i=t)$. Set $p=p_t$, $p'=p_{1-t}$, and
$r:=\PP(T_{-i}=t\mid T_i=t)$, $s:=1-r>0$. Under the peer's
reports $(1-T_{-i},1-p_{T_{-i}})$, the probability of a peer answer
$1$ is $q=1-p$. The best-response calculation of
Lemma~\ref{lem:support} gives, up to a constant $c_0$ independent
of the reported answer,
\[
\begin{aligned}
\max_z U_t(j,z)&=c_0+(1+\kappa r_j)G(z_j^*)
-\kappa\E[\one_{\{K=j\}}G(U)\mid T_i=t],\\
z_j^*&=\frac{q+\kappa\E[U\one_{\{K=j\}}\mid T_i=t]}{1+\kappa r_j}.
\end{aligned}
\]
For $j=1-t$, the unique optimal prediction is $1-p$ and the
maximum is $c_0+G(1-p)$. For $j=t$ it is
\[
\begin{aligned}
z_t^*&=\frac{(1-p)+\kappa s(1-p')}{1+\kappa s},\\
\max_z U_t(t,z)&=c_0+(1+\kappa s)G(z_t^*)-\kappa sG(1-p')\\
&<c_0+G(1-p),
\end{aligned}
\]
by strict convexity of $G$ and $p\ne p'$. Thus the report
$(1-t,1-p_t)$ is the unique best response, and the second profile
is indeed a Bayesian equilibrium. In both cases the bonuses and base
payments attain the bound.
 As honesty gives $(C+\lambda M,C+\lambda M)$, any ex-ante Pareto
 improvement contradicts the aggregate bound; an interim Pareto improvement
 implies an ex-ante one because $0<\pi=\PP(T_i=1)<1$. If an equilibrium has a type with
 payoff $-\infty$, it cannot dominate the finite honest-payoff profile.
\end{proof}

\begin{proof}[Proof of Theorem~\ref{thm:garbling}]
 Recall $\pi=\PP(T_i=1)$, $v=\operatorname{Cov}(T_1,T_2)>0$,
 and $M$ from \eqref{eq:M}. Let $\alpha:=\PP(S_i=1\mid T_i=0)$,
 $\delta:=\PP(S_i=1\mid T_i=1)-\alpha$ for the common garbling,
 and $q:=\alpha+\pi\delta\in(0,1)$. Then
 \[
 c:=\operatorname{Cov}(S_1,S_2)=\delta^2v\ge0,\qquad
 M_S=\frac c{q^2+c}+\frac c{(1-q)^2+c}.
 \]
 If $\delta\ne0$, $c>0$, and Lemma~\ref{lem:channel} gives
 $M_S\le M$, with equality only for the common channel
 $S_i=T_i$ or the common channel $S_i=1-T_i$.
 If $\delta=0$, $M_S=0<M$.
 Hence $C+\lambda M_S\le C+\lambda M$, with the stated equality cases.
\end{proof}

\subsection{Proof of Theorem~\ref{P:pair}}\label{app:implementation}
\begin{proof}
 Recall $I_\epsilon=[\epsilon,1-\epsilon]$,
 $\Dp^\epsilon=\{\epsilon\le y<x\le1-\epsilon\}$,
 and $\widehat V_t,a_1,a_0,B,L$ from the explicit construction
preceding Theorem~\ref{P:pair}.
 Fix $(x,y)\in\Dp^\epsilon$ and let $p_1=x$, $p_0=y$. For the two
 normalized payoff functions, \eqref{alphagamma} and the choice of
 $a_1,a_0$ imply
 \begin{align*}
 &(1-y)\widehat V_1(x,y)-(1-x)\widehat V_0(x,y)\\
 &\quad=(1-y)V_1(x,y)-(1-x)V_0(x,y)-(x-y)a_1<0,\\
 &x\widehat V_0(x,y)-y\widehat V_1(x,y)\\
 &\quad=xV_0(x,y)-yV_1(x,y)-(x-y)a_0<0.
 \end{align*}
 Hence
 \begin{equation}\label{eq:implementation-weakIC}
 \frac{1-y}{1-x}\widehat V_1(x,y)
 <\widehat V_0(x,y)
 <\frac{y}{x}\widehat V_1(x,y).
 \end{equation}

 Let $U_t(j,z)$ be the interim expected payment of type $t$ from
 report $(j,z)$ against an honest peer, under
 \eqref{eq:implementation-mechanism}.
 Against an honest peer, the terms $a_1$ and $a_0$ contribute
 $H_t:=p_ta_1+(1-p_t)a_0$, independently of one's report.
 The honest expected payments are therefore
 \begin{align*}
 U_1(1,x)&=H_1+(1-x)\frac{\widehat V_1(x,y)}{1-x}
          =V_1(x,y),\\
 U_0(0,y)&=H_0+y\frac{\widehat V_0(x,y)}{y}
          =V_0(x,y).
 \end{align*}
 For $(j,z)\in\{0,1\}\times I_\epsilon$,
 \begin{align*}
 j=1,\ z\ne x:\quad
 U_t(1,z)-H_t
 &= (1-p_t)\frac{\widehat V_1(z,y)}{1-z}-p_tL
 \le\frac{B}{\epsilon}-\epsilon L<-B,\\
 j=0,\ z\ne y:\quad
 U_t(0,z)-H_t
 &= p_t\frac{\widehat V_0(x,z)}{z}-(1-p_t)L
 \le\frac{B}{\epsilon}-\epsilon L<-B.
 \end{align*}
 Since $U_t(t,p_t)-H_t=\widehat V_t(x,y)\ge-B$, all these deviations
 are strictly unprofitable. The two remaining nontruthful reports satisfy
 \begin{align*}
 U_1(0,y)-H_1&=\frac{x}{y}\widehat V_0(x,y)
                     <\widehat V_1(x,y),\\
 U_0(1,x)-H_0&=\frac{1-y}{1-x}\widehat V_1(x,y)
                     <\widehat V_0(x,y),
 \end{align*}
 by \eqref{eq:implementation-weakIC}. Thus honesty is strictly
 optimal for each type over $\mathcal R_\epsilon$, establishing strict IC and the
 stated interim payments. The construction uses only boundedness and
 imposes no continuity on the implementing functions.
\end{proof}

\begin{proof}[Proof of Proposition~\ref{prop:impl-necessary}]
Let $A(x):=f_{11}(x,x)$ and $d(y):=f_{00}(y,y)$.
The honest type-$1$ payment is
$V_1=xA(x)+(1-x)f_{10}(x,y)$. IC for type $0$ against
$(1,x)$ implies
\[
\begin{aligned}
V_0(x,y)&\ge yA(x)+(1-y)f_{10}(x,y),\\
(1-x)V_0(x,y)&\ge (1-y)V_1(x,y)-(x-y)A(x),\\
A(x)&\ge\frac{(1-y)V_1(x,y)-(1-x)V_0(x,y)}{x-y}.
\end{aligned}
\]
Similarly, $V_0=yf_{01}(y,x)+(1-y)d(y)$ and IC for type
$1$ against $(0,y)$ yield
\[
yV_1(x,y)\ge xV_0(x,y)-(x-y)d(y),\qquad
d(y)\ge\frac{xV_0(x,y)-yV_1(x,y)}{x-y}.
\]
Taking the appropriate suprema yields the claim.
\end{proof}

\end{document}